\documentclass{aa}  
\usepackage{graphicx}
\usepackage{natbib}
\usepackage{txfonts}
\usepackage{lscape}
\usepackage{color}
\definecolor{comment}{rgb}{1,0,0}

\definecolor{tbc}{rgb}{1,0,0}

\definecolor{acc}{rgb}{0.3,0.7,0.7}
\definecolor{old}{rgb}{0,0.7,0}

\definecolor{ntxt}{rgb}{0,0.2,0.8}

\newcommand{\kms}{km\,s$^{-1}$}

\usepackage{amstext}

\begin{document}

\title{Ensemble asteroseismology of pulsating B-type stars in NGC\,6910\thanks{The photometry presented in Sect.~\ref{PhoObs_ch} is only available in electronic form at the CDS via anonymous ftp to cdsarc.u-strasbg.fr (130.79.128.5) or via http://cdsweb.u-strasbg.fr/cgi-bin/qcat?J/A+A/}}
\author{D.~Mo\'zdzierski\inst{1}, 
A.~Pigulski\inst{1}, 
Z.~Ko{\l}aczkowski\inst{1}\fnmsep\inst{2}, 
G.~Michalska\inst{1}, 
G.~Kopacki\inst{1}, 
F. ~Carrier\inst{3},
P.~Walczak\inst{1}, 
A.~Narwid\inst{1}, 
M.~St\k{e}\'slicki\inst{1}\fnmsep\inst{4}, 
J.-N.~Fu\inst{5}, 
X.-J.~Jiang\inst{6},
Ch.~Zhang\inst{5}, 
J.~Jackiewicz\inst{7}, 
J.~Telting\inst{8}, 
T.~Morel\inst{3}\fnmsep\inst{9}, 
S.~Saesen\inst{3}\fnmsep\inst{10},
E.~Zahajkiewicz\inst{1}, 
P.~Bru\'s\inst{1},
P.~\'Sr\'{o}dka\inst{1}, 
M.~Vu\v{c}kovi\'c\inst{3}\fnmsep\inst{11},
T.~Verhoelst\inst{3}\fnmsep\inst{12},
V.~Van Helshoecht\inst{3},
K.~Lefever\inst{3}\fnmsep\inst{12},
C.~Gielen\inst{3},
L.~Decin\inst{3},
J.~Vanautgaerden\inst{3},
\and 
C.~Aerts\inst{3}\fnmsep\inst{13}
}

\institute{Instytut Astronomiczny, Uniwersytet Wroc{\l}awski, Kopernika 11, 51-622 Wroc{\l}aw, Poland\\ %inst 1
\email{mozdzierski@astro.uni.wroc.pl}
\and Nicolaus Copernicus Astronomical Center, Polish Academy of Sciences, Bartycka 18, 00-716 Warszawa, Poland %inst 2
\and Instituut voor Sterrenkunde, KU Leuven, Celestijnenlaan 200D, 3001 Leuven, Belgium%inst 3
\and Space Research Centre, Polish Academy of Sciences, Kopernika 11, 51-622 Wroc{\l}aw, Poland %inst 4
\and Department of Astronomy, Beijing Normal University, 100875 Beijing, China%inst 5
\and National Astronomical Observatories, Chinese Academy of Sciences, 20A Datun Road, Chaoyang District, 100012 Beijing, China %inst 6
\and Department of Astronomy, New Mexico State University, Las Cruces, NM 88003, USA %inst 7
\and Nordic Optical Telescope, Rambla Jos\'e Ana Fern\'andez P\'erez 7, 38711 San Antonio, Bre\~{n}a Baja,
         Santa Cruz de Tenerife, Spain %inst 8
\and Space sciences, Technologies and Astrophysics Research (STAR) Institute, Universit\'e de Li\`ege, Quartier Agora, All\'ee du 6 Ao\^ut 19c, B\^at.~B5C, B4000-Li\`ege, Belgium%inst 9
\and D\'epartement d’Astronomie, Universit\'e de Gen\`eve, Chemin des Maillettes 51, 1290 Versoix, Switzerland%inst 11
\and Instituto de F\'{\i}sica y Astronom\'{\i}a, Universidad de Valpara\'{\i}so, Casilla 5030, Valpara\'{\i}so, Chile %inst 12
\and Royal Belgian Institute for Space Aeronomy, Ringlaan 3, 1180 Brussels, Belgium % inst 13
\and Department of Astrophysics, IMAPP, Radboud University Nijmegen, 6500 GL Nijmegen, The Netherlands%inst 13
}
\date{Received ...; accepted ...}

% \abstract{}{}{}{}{} 
% 5 {} token are mandatory
\abstract
% context heading (optional)
%leave it empty if necessary  
{Asteroseismology offers the possibility of probing stellar interiors and testing evolutionary and seismic models. Precise photometry and spectroscopy obtained during multi-site campaigns on young open clusters allows discovering rich samples of pulsating stars and using them in a simultaneous seismic modelling called ensemble asteroseismology. The aim of this study is to obtain the age of the open cluster NGC\,6910 by means of ensemble asteroseismology of the early-type pulsating members, to derive their stellar parameters, and to classify the excited modes. We used time-series analysis, performed photometric and spectroscopic mode identification, and calculated grids of evolutionary and seismic models to apply the procedure of ensemble asteroseismology for nine pulsating members of NGC\,6910. With two iterations of the procedure of ensemble asteroseismology applied to nine pulsating stars we derived an age of 10.6$^{+0.9}_{-0.8}$~Myr for NGC\,6910. We also identified the degree $l$ for 8 of 37 modes detected in these stars and classified all modes in terms of $p$, $g$, and mixed-mode pulsations. Of the nine pulsating stars examined in the paper, eight are $\beta$~Cep stars, including three that are hybrid $\beta$~Cep and slowly pulsating B-type (SPB) pulsators, and one is an SPB star. Interestingly, the least massive $\beta$~Cep star, NGC\,6910-38, has a mass of about 5.6\,M$_\odot$. The present theory does not predict unstable $p$ modes in B-type stars with such a low mass. The $g$ modes with relatively high frequencies ($>3.5$\,d$^{-1}$), observed in three members of the cluster, are also stable according to seismic modelling. Both findings pose a challenge for theoretical calculations and prompt a revision of the opacities. The procedure of ensemble asteroseismology was found to be successful for NGC\,6910 and $\chi$~Per on the basis of pulsating B-type stars and can therefore be applied to other young open clusters that are rich in such stars.}
\authorrunning{Mo\'zdzierski et al.}
\maketitle
   
%
%-------------------------------------------------------------------
\section{Introduction}
In the past decades, asteroseismology has been proven to be a fruitful method for probing interiors of pulsating stars, both at the main sequence \citep[e.g.][and references therein]{2017A&A...598A..74P,2018MNRAS.478.2243S,2018pas8.conf...64D} and in the advanced phases of stellar evolution \citep{2013ARA&A..51..353C,2016frap.confE..29D}. By revealing profiles of internal rotation, asteroseismology may play a crucial role in explaining so far not fully understood processes of angular momentum transport during stellar evolution for stars with convective cores \citep{2019ARA&A..57...35A}. Two main factors make asteroseismology a successful technique. The first is a large number of identified pulsation modes. They can be obtained from analysing precise photometry delivered by space missions or intensive ground-based campaigns and surveys. Space missions like Microvariability and Oscillation of Stars (MOST), Convection Rotation and Planetary Transits (CoRoT), Kepler, BRIte Target Explorer (BRITE), and recently, Transiting Exoplanet Survey Satellite (TESS), are particularly good sources of precise photometry. The other factor is a good knowledge of the stellar parameters for the modelled objects. In this context, open clusters appear to be particularly suitable for asteroseismology of their members. The members of an open cluster can be assumed to be coeval and have the same initial chemical composition and distance. Therefore, any information that can be gained from the study of a cluster member or the whole cluster may serve as a constraint for seismic modelling of the member stars. An example is presented by \cite{Fox2006}, who performed seismic modelling of six $\delta$~Sct-type stars in the Pleiades assuming their coevality and the same distance and age. This approach is called ensemble asteroseismology (hereafter EnsA for short), that is,~simultaneous seismic modelling of more than one star, which uses constraints on parameters that result from the membership of these stars to a distinct stellar system. An anticipated output of the EnsA is a consistent picture of pulsations among cluster members on the one hand and determination of global cluster parameters on the other. 

In general, any type of pulsator or even several different groups of pulsating stars simultaneously can be used to conduct EnsA for a given stellar system. The dependence of the periods of pulsating stars on the global stellar parameters leads to the occurrence of useful relations between them, such as the~period-luminosity relation, which is best known for radial modes in classical Cepheids. Relations of this type have been indicated for different pulsating stars and stellar systems. For example, \citet{Bal1997} found relations between mass and pulsation frequency for $\beta$~Cep-type stars in three young open clusters that are rich in these stars, NGC\,3293, NGC\,4755, and NGC\,6231, whose slopes depended on age. The authors subsequently used these slopes to derive the cluster ages.

The availability of the large amount of excellent-quality data from space telescopes, especially Kepler \citep[][and references therein]{Bor2016}, led to the application of the EnsA to red giants in open clusters, especially the three oldest open clusters located in the Kepler field, NGC\,6791, NGC\,6811, and NGC\,6819. \citet{Stel2010} used Kepler data to analyse light curves of red giants in the old open cluster NGC\,6819. Using their solar-like oscillations, the authors were able to conclude on the membership of red giants in the field of this cluster. The work was later extended by \cite{2011ApJ...739...13S} to three open clusters. \citet{Wu2014b} and \citet{Wu2014a} used red giants in NGC\,6819 and NGC\,6791 to determine distance moduli to these clusters and their parameters (including masses) using new scaling relations they derived from the characteristics of solar-like oscillations. \citet{Sand2016} showed the comparison of the age and distance determinations of NGC\,6811 using both red giants and $\delta$ Sct stars in the cluster. Ensembles of red giants in NGC\,6811 and NGC\,6791 were also used for testing their stellar structure, physics, and evolutionary stage \citep{Handb2017,Boss2017}.

The stellar system we focus on here is the young open cluster NGC\,6910. After the discovery that the cluster is rich in $\beta$~Cep stars \citep{Zibi2004}, a joint photometric campaign on NGC\,6910, h and $\chi$ Per, and NGC\,3293 was organized \citep{2008CoAst.157..338M,2008JPhCS.118a2071S,Pigu2008,2007CoAst.150..193H,2008CoAst.157..315H}. The campaign was conducted in 2005\,--\,2007. It led to the discovery of new $\beta$~Cep stars in all clusters, but NGC\,3293 and showed that they have a potential for the EnsA \citep{Pigu2008,Saes2010b}. In particular, $\beta$~Cep stars in NGC\,6910 revealed a clear dependence of their frequencies on brightness. The full results of the campaign on $\chi$~Per were published by \cite{2010A&A...515A..16S}. These results were subsequently used by \cite{Saes2013} to constrain the age of $\chi$~Per by comparing the observed frequencies of five $\beta$~Cep stars with the frequencies calculated for a large set of theoretical models, imposing their coevality and the same initial chemical composition.

We here apply EnsA to pulsating, mostly $\beta$~Cep-type, stars in NGC\,6910. We first introduce the cluster itself (Sect.~\ref{cluster}). Then, the photometric (Sect.~\ref{PhoObs_ch}) and spectroscopic (Sect.~\ref{SpeObs_ch}) data we use are described. Next, the procedure of the EnsA and its results are presented (Sect.~\ref{EnsA_ch}) and discussed (Sect.~\ref{discussion}).

\section{Open cluster NGC\,6910}\label{cluster}
NGC\,6910 (C 2021+406, OCl 181) is a young open cluster located about 1$\degr$ north of $\gamma$~Cygni (Sadr) in the region of Cyg\,OB9 association \citep[e.g.][]{1992A&AS...94..211G}. The discrepancy of the distances of the cluster and Cyg OB9 indicates that the two systems are not physically related \citep{Com2012}, however. NGC\,6910 has been the subject of frequent photometric studies that were summarized by \citet{Zibi2004}. The cluster distance is estimated to be 1.1\,--\,1.5~kpc \citep{2005A&A...438.1163K,2013A&A...558A..53K}, which places it behind the Great Cygnus Rift. In result, the reddening across the cluster, as determined in its central parts by \cite{Zibi2004}, is variable and amounts to 1.25\,--\,1.7~mag in terms of $E(B-V)$ colour excess. \cite{Zibi2004} determined the age of NGC\,6910 by isochrone fitting on $6\pm2$~Myr. 

The first variability survey in NGC\,6910 was carried out by \citet{Zibi2004}, who found six variable stars in the cluster, including four $\beta$~Cep-type stars: NGC\,6910-14, 16,  18, and 27. The numbering of stars in the cluster follows the one used in the WEBDA database\footnote{https://webda.physics.muni.cz}. We use it throughout the paper. The discovery placed NGC\,6910 among a handful of young open clusters that are rich in $\beta$~Cep-type stars and prompted us to carry out the campaign described in the Introduction. The campaign resulted in the discovery of five candidates for $\beta$~Cep stars: NGC\,6910-25, 34, 36, 38, and 41, each showing at least one frequency higher than 5\,d$^{-1}$. For the known $\beta$~Cep stars, many new modes were detected. For example, a dozen modes were found in NGC\,6910-18 \citep{Pigu2008}. In the present paper we deal only with the nine stars listed above. We call them hereafter `program stars'. Their frequency spectra were used to carry out EnsA in NGC\,6910. The full variability survey resulting from the photometry obtained during the 2005\,--\,2007 campaign is beyond the scope of this paper and will be published separately (Mo\'zdzierski et al., in preparation). A very preliminary result of this survey was presented by \citet{Moz2018}: more than 120 variable stars in the field of NGC\,6910 were found.

\section{Photometry}\label{PhoObs_ch}
\subsection{Observations and calibration}
The 2005\,--\,2007 campaign on NGC\,6910 resulted in CCD photometry of NGC\,6910 of different quality and quantity carried out in 11 observatories. For the purpose of this paper, we chose only the longest and best-quality data obtained in 3 of the 11 sites: our own observatory (Bia{\l}k\'ow, Poland), Xinglong Observatory (China), and the Observatorio del Roque de los Muchachos (ORM, La Palma, Spain).

\begin{figure}
\centering
\includegraphics[width=0.9\columnwidth]{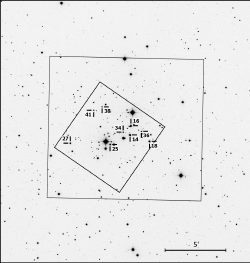}
\caption{Comparison of the fields in NGC\,6910 covered by the Xinglong (the largest), Bia{\l}k\'ow (intermediate), and ORM (the smallest field) observations. The nine program stars are labelled with their numbers from WEBDA. North is up, and east is to the left.}
\label{FoVs}
\end{figure}

The data in Bia{\l}k\'ow were gathered with the 60cm Cassegrain telescope equipped with Andor Tech.~DW432-BV back-illuminated CCD camera covering a 13$^\prime$\,$\times$\,12$^\prime$ field of view in the $B$, $V$, and $I_{\rm C}$ passbands of the Johnson-Kron-Cousins photometric system. Observations in Xinglong were carried out with the 50cm reflecting telescope equipped with a VersArray 1300B CCD camera covering 22$^\prime$\,$\times$\,22$^\prime$ field of view and Johnson $V$ filter. In the ORM, the Mercator 1.2m telescope equipped with the EEV 40-42 (Merope) CCD camera covering a 6$^\prime\!\!.5$\,$\times$\,6$^\prime\!\!.5$ field of view was used. The observations were obtained in the Geneva $U$ (hereafter $U_{\rm G}$), Johnson $B$ and $V$, and Cousins $I_{\rm C}$ passbands, but for the purpose of the present paper, we used only the photometry obtained in the $U_{\rm G}$ filter. In total, the Bia{\l}k\'ow and Xinglong data we used consist of a set of about 3800 CCD frames in the $B$, 19\,800 in the $V$, and 5800 in the $I_{\rm C}$ passbands collected during 138 observing nights between August 2005 and October 2007. We also used about 1350 CCD frames obtained through the $U_{\rm G}$ filter in ORM during 116 nights between April 2005 and August 2007. A comparison of the observed fields in the three observatories is shown in Fig.~\ref{FoVs}.

The Bia{\l}k\'{o}w and Xinglong observations were calibrated in the standard way, which included dark and bias subtraction and flat-field correction. The ORM data required more complicated reduction as the flat-fields were scarce and for the data obtained between July and August 2006 suffered from the long-term changes due to the installation of a heat shield in the nitrogen dewar. To solve the problem, we interpolated scaled flat-fields by six-order polynomials to the observation times of the science frames. The procedure followed the one applied by \cite{2010A&A...515A..16S} to the $\chi$~Per data, which were obtained with the same equipment.

Supplementary photometry of NGC\,6910 was obtained in Bia{\l}k\'ow during 21 observing nights between July and October 2013. The purpose of these observations was to acquire simultaneous photometry with the time-series spectroscopy of some $\beta$~Cep stars in NGC\,6910 at Apache Point Observatory (APO, Sect.\,\ref{spec-obs}). In total, we obtained 460 CCD frames in the $B$, 2500 in the $V$, and 1100 in the $I_{\rm C}$ filters. They were calibrated in the same way as the data from the campaign.

\subsection{Reduction and time-series analysis}
The calibrated CCD frames were reduced with the {\sc Daophot II} package \citep{Stet1987}. In total, the list of detected objects included 1978 stars. The result was both profile (point spread function, PSF) and aperture photometry. The latter was obtained in the following way. First, all stars were subtracted from a given frame through profile photometry. Then, stars were added to the frame one by one, and we obtained the aperture photometry with apertures scaled with the seeing. In general, the aperture photometry showed smaller scatter than the profile photometry for the brightest stars, but this depended on crowding. For six program stars (NGC\,6910-14, 16, 18, 27, 36, and 41) the aperture photometry from Bia{\l}k\'ow and Xinglong was combined. For NGC\,6910-25, the profile photometry from Bia{\l}k\'ow was combined with the aperture photometry from Xinglong. For NGC\,6910-34 and 38 we used data only from Bia{\l}k\'ow, profile photometry for the former, and aperture photometry for the latter. Finally, for the $U_{\rm G}$ ORM data, we used aperture photometry for all program stars except for NGC\,6910-36, for which the profile photometry was chosen.
\begin{figure}
\centering
\includegraphics[width=\columnwidth]{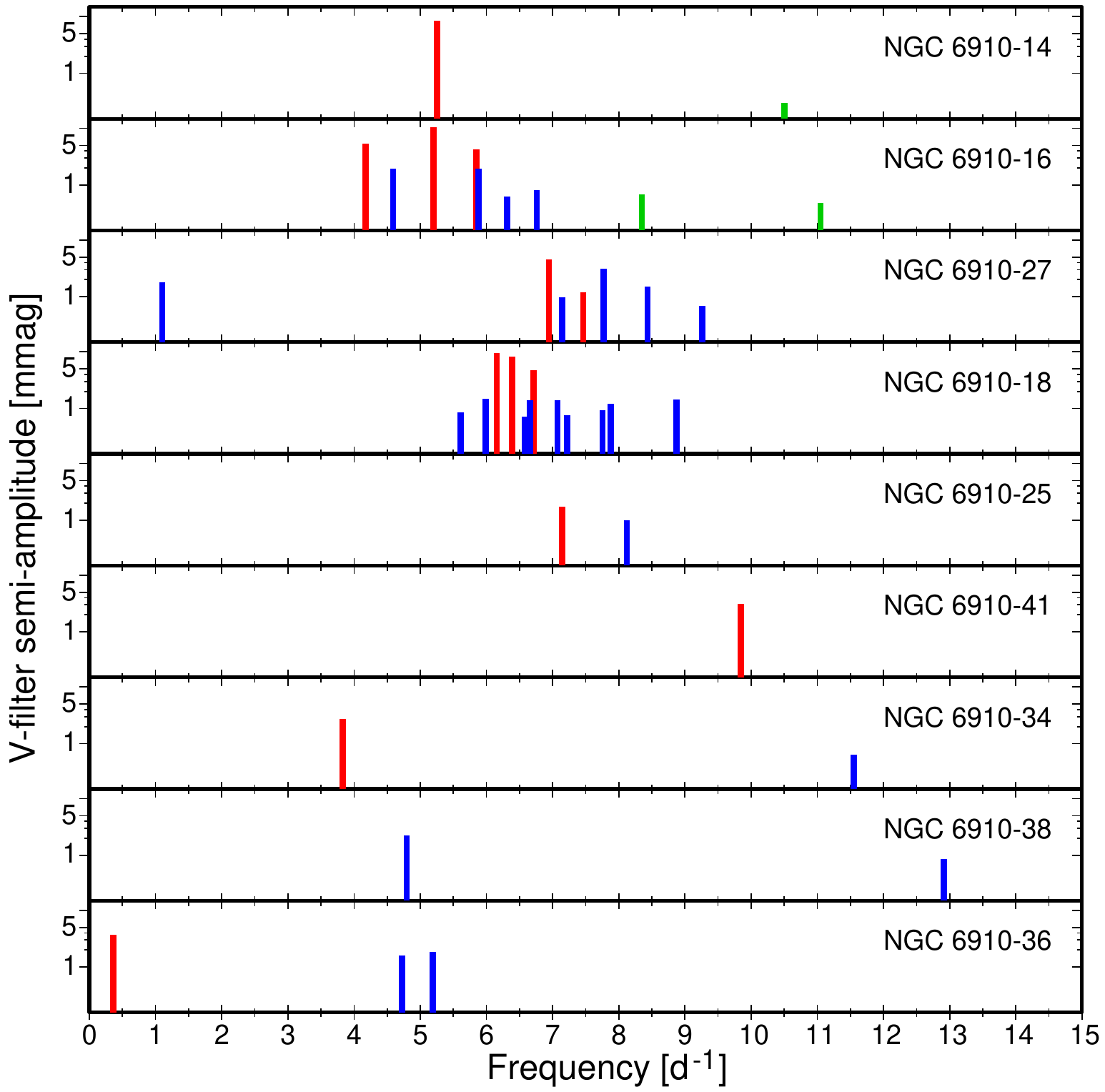}
\caption{Schematic frequency spectra of the nine program stars. The intrinsic modes are shown as red lines, if detected both in the 2005\,--\,2007 and 2013 $V$-filter data, or blue lines, if detected only in the 2005\,--\,2007 data. Detected harmonics and combination frequencies are shown with green lines.}
\label{9fs}
\end{figure}

The campaign photometry was the subject of the subsequent time-series analysis. Because most data were obtained in the Johnson $V$ passband, we used this photometry to reveal frequency spectra of the nine program stars. The spectra are shown in Fig.\,\ref{9fs}. In total, we detected 40 terms including 37 intrinsic frequencies, two harmonics, and one combination frequency. Two stars, NGC\,6910-14 and 41, show only a single mode, although in the former, the harmonic of the mode was also detected. The remaining seven stars are multi-periodic, which is a strong argument in favour of pulsations as the cause of their variability. Thus, we hereafter attribute the intrinsic frequencies to pulsation modes. The richest spectrum was confirmed for NGC\,6910-18, in which 12 modes were found. In two stars, NGC\,6910-27 and 36, modes with frequencies below 1.5~d$^{-1}$ were detected. Given their frequencies, they have to be $g$ modes. To distinguish whether the other modes are $p\text{}$, $g\text{,}$ or mixed $p/g$ modes requires that they are identified. This question is addressed in Sect.\,\ref{EnsA_ch}.

The number of observations in the Geneva $U_{\rm G}$, Johnson $B$ and Cousins $I_{\rm C}$ data was lower than in the Johnson $V$ band, and consequently the number of terms detected was also lower. For all nine program stars we detected in total 17 terms in the $U_{\rm G}$, 39 terms in the $B$, and 34 terms in the $I_{\rm C}$ data. The results of the sine-curve fits to the $U_{\rm G}BVI_{\rm C}$ data for the program stars are presented in Table \ref{1stPerPar} (Appendix \ref{sinefits}). 
\begin{figure}
\centering
\includegraphics[width=0.8\columnwidth]{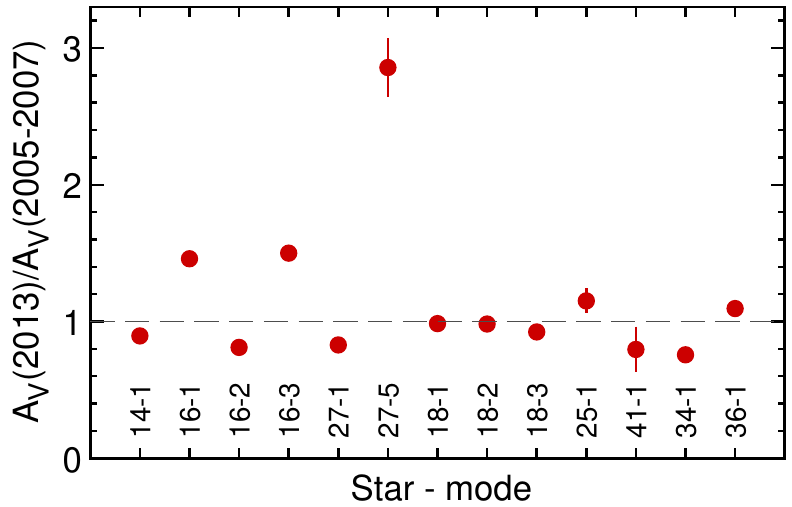}
\caption{Ratios $A_V(2013)/A_V(2005-2007)$ of the $V$-filter amplitudes for modes detected in both campaigns. The modes are identified with star and mode number, according to the identification in Table \ref{1stPerPar}. For example, the mode labelled 18-2 is the $f_2$ mode in NGC\,6910-18.}
\label{amp-comp}
\end{figure}

The $VI_{\rm C}$ data from the 2013 observations were analyzed in the same way as the campaign data, and in total, we detected 13 terms in the $B$, 12 in the $V$, and 8  in the $I_{\rm C}$ data (Table \ref{2ndPerPar}). No variability of NGC\,6910-38 was detected in 2013 data. Some modes between the 2005\,--\,2007 campaign and 2013 show significant changes in amplitude. This is illustrated in Fig.\,\ref{amp-comp}, in which the ratios of amplitudes, $A_V(2013)/A_V(2005-2007)$, are shown. For three modes, $f_1$ and $f_3$ in NGC\,6910-16, and for $f_5$ in NGC\,6910-27, the amplitudes increased by over 30\%. The most remarkable is an almost threefold increase of the $V$-filter amplitude for $f_5$ in NGC\,6910-27, from 1.19\,$\pm$\,05~mmag during the 2005\,--\,2007 campaign to 3.40\,$\pm$\,0.21~mmag in 2013. This is an illustration of the intrinsic long-term amplitude changes in $\beta$~Cep stars, which are known in many members of this class \citep[e.g.][]{Jerzyk2015}.

\subsection{$VI_{\rm C}$ photometry}
\begin{figure}
\centering
\includegraphics[width=\columnwidth]{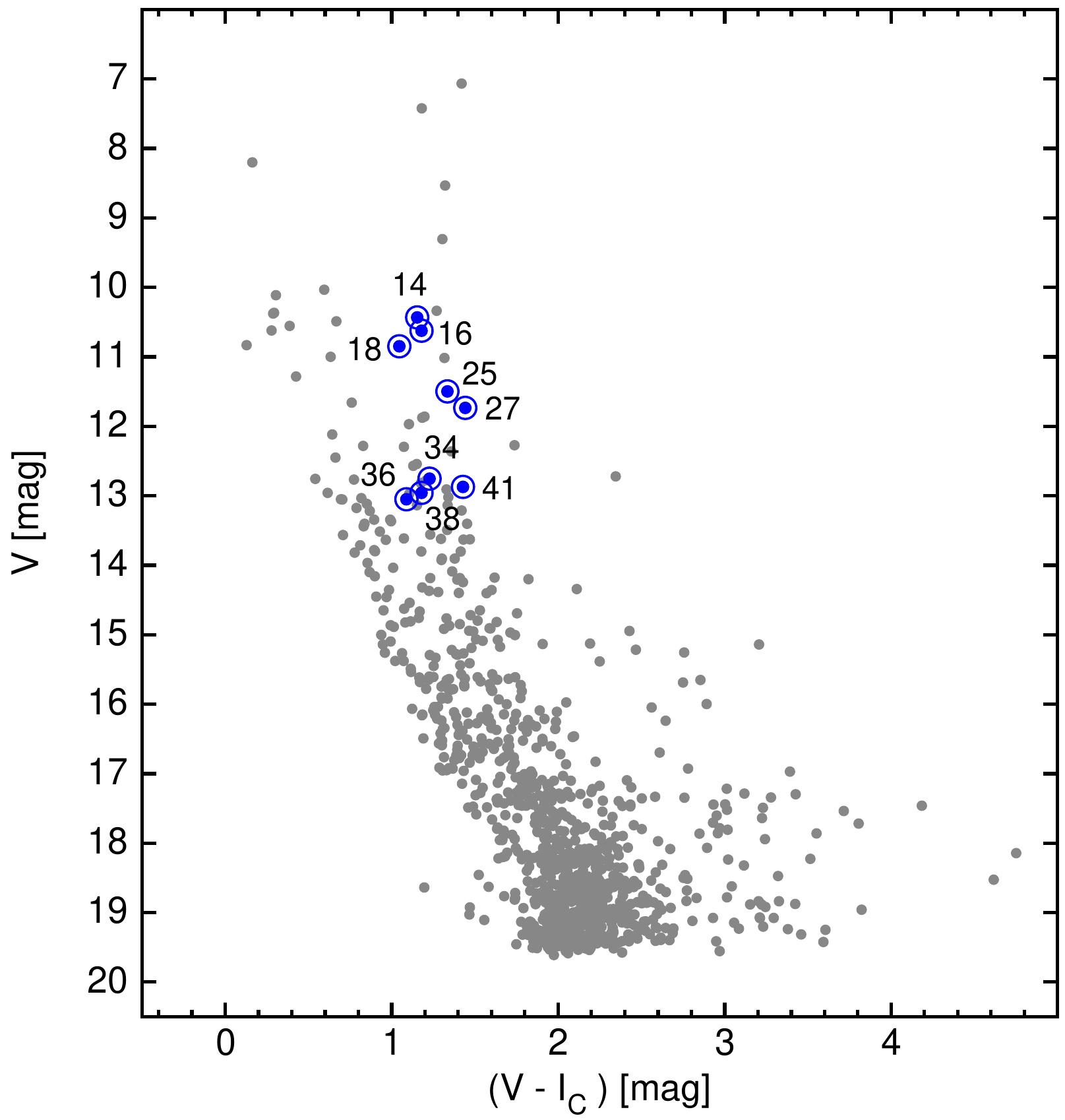}
\caption{Colour-magnitude diagram for the field of NGC\,6910 observed in Bia{\l}k\'ow in 2013. Program stars are marked with blue encircled symbols and are labelled with their WEBDA numbers.}
\label{CMD}
\end{figure}

The instrumental $VI_{\rm C}$ magnitudes obtained during 2013 observations were transformed into the standard system. The mean instrumental magnitudes were tied to the photometry of \citet{Zibi2004} by means of the following transformation equations:
\begin{equation}\label{VK04}
 V=v-(0.0802\pm 0.0064)\times(v-i_{\rm{C}})+(10.7647\pm 0.0039),
\end{equation}
\begin{equation}\label{VIK04}
 (V-I_{\rm{C}})=(0.9181\pm 0.0047)\times(v-i_{\rm{C}})+(1.0088\pm 0.0029),
\end{equation}
where the lowercase letters denote instrumental and uppercase letters the standard magnitudes. The transformation equations were obtained by means of the least-squares method using data for 112 stars. The standard deviations for Eqs.~(\ref{VK04}) and (\ref{VIK04}) are equal to 0.032~mag and 0.024~mag, respectively. The $VI_{\rm C}$ photometry of the observed field is shown in the $(V-I_{\rm C})$ versus $V$ colour-magnitude diagram (hereafter CMD; Fig.\,\ref{CMD}). Owing to the high and variable reddening across the cluster, the brightest members are not the bluest objects in the CMD. They form a smeared sequence located to the right of the sequence of mostly foreground late-type main-sequence stars. In the lower part of CMD, cluster and field stars overlap.

\section{Spectroscopy} \label{SpeObs_ch}
\subsection{Observations}\label{spec-obs}
Out of nine program stars, the five brightest were observed spectroscopically either during the campaign or later, in 2013. The campaign data were obtained at two observatories, at the ORM at La Palma, Spain, and at the Observatoire de Haute-Provence (OHP) in France. The ORM observations were made with the 2.56m Nordic Optical Telescope (NOT) with the FIES \'{e}chelle spectrograph. During these observations, FIES worked in the medium-resolution mode with a resolving power of $R=46\,000$, and a fibre diameter of $1.\!\!^{\prime\prime}3$. The spectra range between 370 and 730~nm spread over 77 orders. The OHP observations were carried out with the 1.93m telescope equipped with the SOPHIE \'{e}chelle spectrograph covering 39 orders in the spectral range between 387.2 and 694.3~nm. The resolving power of this spectrograph equals 75\,000 and its fibre diameter corresponds to $3^{\prime\prime}$. 

In 2013, spectroscopic observations were made at the APO, New Mexico, USA, with the Astrophysical Research Consortium (ARC) 3.5-meter telescope equipped with the \'{e}chelle spectrograph ARCES (ARC Echelle Spectrograph). ARCES spectra consist of 120 orders covering a spectral range of 320\,--\,1000~nm and have a resolving power of about 31\,500. We used a slit that covers a $1.\!\!^{\prime\prime}6\times3.\!\!^{\prime\prime}2$ field of view.

The journal of spectroscopic observations is given in Table \ref{SpecList}. In total, we obtained 87 spectra for NGC\,6910-14, 5 spectra for NGC\,6910-16, 44 spectra for NGC\,6910-18, and a single spectrum for NGC\,6910-25 and 27. All spectra were reduced using standard IRAF\footnote{IRAF is distributed by the National Optical Astronomy Observatory, which is operated by the Association of Universities for Research in Astronomy, Inc., under the cooperative agreement with the National Science Foundation.} routines, instructions provided in the ARCES guide\footnote{http://astronomy.nmsu.edu:8000/apo-wiki/wiki/ARCES\#reduction}, FIEStool\footnote{http://www.not.iac.es/instruments/fies/fiestool/}, the package dedicated for FIES spectra, and guide for reducing SOPHIE spectra\footnote{http://www.obs-hp.fr/guide/sophie/sophie-eng.shtml.}.
\begin{table}
\caption{Journal of spectroscopic observations for program stars in NGC\,6910.}
\label{SpecList}
\centering\small
\begin{tabular}{lcl}
\hline\hline\noalign{\smallskip}
Date & Site & Stars observed (number of spectra)\\
\noalign{\smallskip}\hline\noalign{\smallskip}
2007 Sep 24 & ORM & NGC\,6910-18 (9) \\
2007 Sep 25 & ORM & NGC\,6910-14 (9)\\
2007 Sep 26 & ORM & NGC\,6910-14 (6)\\
2007 Sep 27 & ORM & NGC\,6910-14 (6), 16 (2)\\
2007 Sep 28 & ORM & NGC\,6910-16 (2), 25 (1), 27 (1)\\
2007 Oct 13 & OHP & NGC\,6910-14 (12)\\
2007 Oct 14 & OHP & NGC\,6910-14 (11)\\
2007 Oct 15 & OHP & NGC\,6910-14 (8)\\
2007 Oct 16 & OHP & NGC\,6910-14 (7)\\
2007 Oct 17 & OHP & NGC\,6910-14 (7)\\
2007 Oct 18 & OHP & NGC\,6910-14 (7)\\
2007 Oct 19 & OHP & NGC\,6910-14 (7)\\
2007 Oct 20 & OHP &NGC\,6910-14 (6)\\
2013 June 6 & APO & NGC\,6910-14 (1), 18 (1)\\
2013 July 13 & APO & NGC\,6910-16 (1), 18 (2)\\
2013 Aug 14 & APO & NGC\,6910-18 (4)\\
2013 Sep 2 & APO & NGC\,6910-18 (9)\\
2013 Sep 29 & APO & NGC\,6910-18 (22)\\
\noalign{\smallskip}\hline
\end{tabular}
\end{table}

\subsection{Determination of stellar parameters} \label{AtmPar_ch}
Because the APO and NOT spectra have the best quality, they were used to derive stellar parameters of the five brightest program stars. Prior to this operation, the spectra for three stars with more than one spectrum were averaged. This operation increased the signal-to-noise ratio (S$/$N) and reduced the influence of pulsations on line profiles. Given the number of the obtained spectra, this was the most effective for NGC\,6910-14 and 18. Because the spectra were averaged, the overall line broadening includes both rotational and pulsational broadening and hence is an upper limit of the projected rotational velocity. Based on the results presented in Sect.\,\ref{SpeId_ch}, the contribution of the latter broadening is expected to be smaller than $\sim$20\% for all program stars, however.

The stellar parameters were determined in two steps. In the first step, we fitted interpolated model spectra to the selected parts of the observed spectra by means of non-linear least squares. The selected parts of spectra included the lines of \ion{He}{i}, \ion{He}{ii}, \ion{O}{ii}, \ion{N}{ii}, \ion{S}{iii}, \ion{C}{ii}, \ion{Fe}{ii}, \ion{Fe}{iii}, and \ion{Si}{iv}. We avoided hydrogen lines because the normalisation and correction for blaze function affected their wings. The model spectra were taken from NLTE BSTAR2006 grid of models \citep{Lanz2007}. The parameters derived in these fits were the following: effective temperature, $T_{\rm eff}$, surface gravity, $\log g$, metallicity, $Z/Z_\odot$, projected rotational velocity, $V_{\rm eq}\sin i$, and average radial velocity, $\langle V_{\rm r}\rangle$. The models were interpolated in three-dimensional space ($T_{\rm eff}$, $\log g$, and $Z/Z_\odot$); the spectra were rotationally broadened by means of the {\tt rotin3} program available at {\sc Synspec} web page\footnote{http://nova.astro.umd.edu/Synspec49/synspec.html.}, rebinned in $\ln\lambda$, and shifted by $\langle V_{\rm r}\rangle$. We expect that all stars, as members of NGC\,6910, have the same metallicity. The weighted average of the values of $Z$ derived from the spectra of three stars (NGC\,6910-14, 16, and 18) is 0.97. In the second step we therefore repeated the fits, this time assuming $Z/Z_\odot$ fixed at 1. The results of these fits are given in Table \ref{atmpartab}.

Moreover, we derived two parameters, $T_{\rm eff}$ and $\log g$, from the calibrations of Str\"{o}mgren photometry of \cite{Cra1977}, \cite{Mar2001}, \cite{Cap2002}, and \cite{Hand2011}, which is available for all nine program stars. The dereddening and calibration of photometry was made by means of the UVBYBETA program\footnote{UVBYBETA computer program was written by T.T.~Moon in 1985 and modified later by R.~Napiwotzki. The program is based on grids presented by \cite{Moon1985}, which can be used  for the determination of $T_{\rm eff}$ and $\log g$.}. Errors of $T_{\rm eff}$ were estimated in the same way as done by \cite{Jerzyk2015} for 16 Lac, that is,~based on the uncertainties of the absolute flux calibrations \citep{Nap1993,Jerzyk1994}. Thus, we adopted a 3\% error for stars with $T_{\rm eff} < 20\,000$ K and a 4\% error for stars with $T_{\rm eff} > 20\,000$ K. Errors of $\log g$ were calculated by propagating errors of the $uvby\beta$ photometry. The results of the determination of $T_{\rm eff}$ and $\log g$ from Str\"{o}mgren photometry for all nine program stars are given in Table \ref{atmpartab}.

For the purpose of EnsA, we needed to decide on the adopted values of the stellar parameters, $Z/Z_\odot$, $T_{\rm eff}$ , and $\log g$, in particular. Following the choice made above, we adopted $Z/Z_\odot = 1$ as representative for cluster members. For the three brightest stars, the effective temperatures were taken from the spectroscopic solution. We decided to adopt more conservative than the formally obtained uncertainties of $T_{\rm eff}$, however, that is, 1000~K. For six cooler stars, we adopted the values obtained from Str\"omgren photometry. For stars with more than one determination, we favoured those obtained from the data taken by \cite{Hand2011} because this photometry was of the best quality. For the remaining stars, we adopted a simple average of the derived values of $T_{\rm eff}$. The adopted values are presented in Table \ref{atmpartab}. Surface gravities are not well constrained from spectroscopic fits because the fitted lines are not very sensitive to $\log g$. (We did not use hydrogen lines in the fits because they are subject to a bias. This bias occurs because a proper normalisation of echelle spectra for wide spectral lines is not possible.) We therefore decided the following. For stars with the best Str\"omgren  photometry, that is, the five brightest stars, we relied on the values derived from this photometry, again giving preference to those derived from \cite{Hand2011} data. For the remaining stars, we adopted $\log g = 4.04\,\pm\,0.20$ for the hotter NGC\,6910-25, 34, and 41, and $\log g = 4.12\,\pm\,0.20$ for the cooler NGC\,6910-36 and 38.

The radial velocities of five brightest program stars (Table \ref{atmpartab}) agree fairly well with the radial velocity of the cluster as given by \cite{2002A&A...389..871D}, $-$31.6~{\kms}, and \cite{2005A&A...438.1163K,2013A&A...558A..53K}, $-$32.8~{\kms}. They are also consistent with  the radial velocity of the emission nebula associated with the cluster, estimated by \cite{Kub2007} for $-$25 to $-$30~{\kms}. For NGC\,6910-18, the two radial velocities obtained at the two epochs separated by six years differ by about 14~{\kms}. This can be evidence of the binarity of the star.
\begin{table*}\small
\centering
\caption{Stellar parameters for the program stars in NGC\,6910.}             
\label{atmpartab}      
\centering          
\begin{tabular}{c c c r c c c c c} 
\hline\hline\noalign{\smallskip}
Star & \multicolumn{4}{c}{Spectroscopy} & \multicolumn{2}{c}{Str\"omgren photometry} & \multicolumn{2}{c}{Adopted}\\
& $T_{\rm eff}$ & $\log g$& $V_{\rm eq}\sin i$ & $\langle V_{\rm r}\rangle$ & $T_{\rm eff}$& $\log g$ & $T_{\rm eff}$ & $\log g$\\
NGC\,6910-&   [K]  &  (cgs)& [{\kms}] & [{\kms}] & [K] &  (cgs)& [K] &  (cgs)\\
\noalign{\smallskip}\hline\noalign{\smallskip}
14    & ${27040^{+210}_{-220}}$ & ${3.77^{+0.031}_{-0.020}}$ & ${148^{+1.5}_{-2.2}}$ & ${-25.3^{+1.2}_{-1.3}}^{\rm \,OHP}$ & $27500\pm1100^{\rm \,H}$ & ${3.77^{+0.22}_{-0.22}}^{\rm \,H}$ & $27000\pm1000$ &  $3.77^{+0.22}_{-0.22}$\\
      &  &  &  & ${-26.5^{+0.7}_{-0.8}}^{\rm \,NOT}$ & $28400\pm1140^{\rm \,Cr}$ & ${3.95^{+0.28}_{-0.25}}^{\rm \,Cr}$ &  & \\
      &  &  &  &  & $29100\pm1160^{\rm \,M}$ & ${4.95^{+0.25}_{-0.25}}^{\rm \,M}$ &  & \\
\noalign{\smallskip}\hline\noalign{\smallskip}
16    & ${25880^{+260}_{-320}}$ & ${3.93^{+0.024}_{-0.030}}$ & ${168^{+1.9}_{-2.3}}$ & ${-28.6^{+1.7}_{-1.6}}^{\rm \,NOT}$ & $25000\pm1000^{\rm \,H}$ & ${4.04^{+0.19}_{-0.22}}^{\rm \,H}$ & $25900\pm1000$ & ${4.04^{+0.19}_{-0.22}}$\\
\noalign{\smallskip}\hline\noalign{\smallskip}
27    & --- & --- & ${115^{+15}_{-13}}$ &  ${-28.0^{+4}_{-4}}^{\rm \,NOT}$ & $25100\pm1000^{\rm \,H}$ & ${4.18^{+0.18}_{-0.21}}^{\rm \,H}$ & $25100\pm1000$ & $4.18^{+0.18}_{-0.21}$\\
      &  &  &  &  & $26500\pm1060^{\rm \,Cr}$ & ${4.16^{+0,24}_{-0,28}}^{\rm \,Cr}$ &  & \\
\noalign{\smallskip}\hline\noalign{\smallskip}
18    & ${24490^{+380}_{-410}}$ & ${3.88^{+0.045}_{-0.050}}$ & ${97^{+1.6}_{-1.2}}$ & ${-34.7^{+0.5}_{-0.4}}^{\rm \,APO}$ & $24800\pm990^{\rm \,H}$ & ${4.05^{+0.19}_{-0.22}}^{\rm \,H}$ & $24500\pm1000$ & ${4.05^{+0.19}_{-0.22}}$\\
      &  &  &  & ${-20.4^{+0.8}_{-0.7}}^{\rm \,NOT}$ & $26000\pm1040^{\rm \,Cr}$ & ${4.36^{+0.23}_{-0.25}}^{\rm \,Cr}$ &  & \\
\noalign{\smallskip}\hline\noalign{\smallskip}
25    & --- & --- & ${153^{+11}_{-13}}$  & ${-34.5^{+7}_{-5}}^{\rm \,NOT}$ & $21900\pm880^{\rm \,H}$ & ${4.04^{+0.20}_{-0.20}}^{\rm \,H}$ & $21900\pm880$ & ${4.04^{+0.20}_{-0.20}}$\\
\noalign{\smallskip}\hline\noalign{\smallskip}
41    & --- & --- & --- & ---  & $20600\pm820^{\rm \,M}$ & ${5.21^{+0.25}_{-0.25}}^{\rm \,M}$ & $20800\pm830$ & ${4.04^{+0.20}_{-0.20}}$\\
      &  &  &  &  & $21300\pm850^{\rm \,Cr}$ & ${4.84^{+0.17}_{-0.20}}^{\rm \,Cr}$ &  & \\
      &  &  &  &  & $20600\pm820^{\rm \,Ca}$ & ${4.65^{+0.30}_{-0.34}}^{\rm \,Ca}$ &  & \\
\noalign{\smallskip}\hline\noalign{\smallskip}
34    & --- & --- & --- & --- & $20800\pm830^{\rm \,M}$ & ${4.94^{+0.25}_{-0.25}}^{\rm \,M}$ & $20700\pm830$ & ${4.04^{+0.20}_{-0.20}}$\\
      &  &  &  &  & $20500\pm820^{\rm \,Cr}$ & ${4.36^{+0.17}_{-0.23}}^{\rm \,Cr}$ &  & \\
\noalign{\smallskip}\hline\noalign{\smallskip}
38    & --- & --- & --- & --- & $17500\pm530^{\rm \,M}$ & ${4.70^{+0.25}_{-0.25}}^{\rm \,M}$ & $18000\pm540$ & $4.12^{+0.14}_{-0.16}$\\
      &  &  &  &  & $18500\pm560^{\rm \,Cr}$ & ${4.12^{+0.14}_{-0.16}}^{\rm \,Cr}$ &  & \\
\noalign{\smallskip}\hline\noalign{\smallskip}
36    & --- & --- & --- & --- & $16500\pm500^{\rm \,M}$ & ${4.80^{+0.25}_{-0.25}}^{\rm \,M}$ & $16950\pm510$ & ${4.12^{+0.14}_{-0.16}}$\\
      &  &  &  &  & $17400\pm520^{\rm \,Cr}$ & ${4.39^{+0.11}_{-0.11}}^{\rm \,Cr}$ &  & \\
\noalign{\smallskip}\hline
\end{tabular}
\tablebib{(H)~\citet{Hand2011}; (Ca) \citet{Cap2002}; (M) \citet{Mar2001}; (Cr) \citet{Cra1977}.}
\end{table*}

\section{Ensemble asteroseismology} \label{EnsA_ch}
The procedure of EnsA adopted in the present paper consists of the following steps:
\begin{enumerate}
\item We identify pulsation modes using both photometric (Sect.\,\ref{PhoId_ch}) and spectroscopic (Sect.\,\ref{SpeId_ch}) observations. Because the precision of the measurements is limited and spectroscopy for faint program stars is lacking, the conclusive result can be obtained only for a limited number of modes in the brightest stars.
\item We define a grid of parameters for evolutionary and pulsation models and use identified modes to place the limits on the cluster age (Sect.~\ref{step-age}).
\item We derive parameters of all program stars using the cluster age as the primary constraint placed on the models. In parallel, we revise the possible identifications and exclude those that do not fit the cluster age and the adopted limits of $T_{\rm eff}$ (Sect.~\ref{step-mid}). In general, the procedure may lead to unambiguous identification of additional modes and/or a further narrowing of cluster age. If this is the case, the last step can be iterated.
\end{enumerate}
The procedure determines the age of the cluster and also the physical parameters of the program stars, including their masses and surface gravities (Sect.\,\ref{params}).

\subsection{Photometric identification of modes} \label{PhoId_ch}
Using photometry, we identified modes by means of the method developed by \citet{Dasz2002,Dasz2005}, in which spherical harmonic degrees $l$ of the observed modes are identified by comparing theoretical and observed amplitude ratios and phase differences. For this purpose, we made use of the $U_{\rm G}BVI_{\rm C}$ photometry from the 2005\,--\,2007 campaign (Sect.\,\ref{PhoObs_ch}, Table \ref{1stPerPar}). The evolutionary models were calculated with the Warsaw-New Jersey code developed by \citet{Pacz1969,Pacz1970}, and modified later by R.\,Sienkiewicz, M.\,Koz{\l}owski, A.\,A.\,Pamyatnykh, and W.\,A.\,Dziembowski \citep{Pamy1998,Dzie2008}. Taking advantage of the membership of program stars, we assumed the AGSS09 solar chemical mixture \citep{Asp2009}, the initial hydrogen abundance $X = 0.72$ and,  in accordance with the results of spectroscopic determination of $Z$ (Sect.\,\ref{AtmPar_ch}), solar metallicity, $Z = Z_{\odot} = 0.014$. All calculations were made using the equation of state and opacity tables from the OPAL project \citep[EOS 2005,][]{Igle1996,Roge2002}.

One of the main problems in seismic modelling is the treatment of stellar rotation. The projected rotational velocities for the five program stars range between 97 and 168~{\kms} (Table \ref{atmpartab}), so that their rotational equatorial velocities can be believed to be far from critical, the more so because emission in H$\alpha$ is not observed in any of them. Because some information on the rotation is required for seismic modelling, we assumed that the rotational velocity is similar in all program stars although the sample is small. Assuming randomly oriented rotational axes, the mean value of $\langle{V_{\rm eq}\sin i}\rangle \approx$ 136~{\kms} for the five program stars translates into a mean equatorial velocity, $\langle{V_{\rm eq}}\rangle$, equal to about 170~{\kms}. It was derived from the equation
\begin{equation}
 \langle{V_{\rm eq}}\rangle=\frac{4}{\pi}\langle{V_{\rm eq}\sin i}\rangle,
\end{equation}
given by \cite{1950ApJ...111..142C}. The relation was recently confirmed by \citet{Sil2014} for stars observed with Kepler mission.

The next parameter that needs to be decided on in stellar modelling is the overshooting parameter, $\alpha_{\rm ov}$, which describes the extent of the convective overshooting from the core in terms of the pressure height scale. The reliable estimates from seismic modelling range between 0 and 0.6 \citep{Aer2015}, but rarely exceed 0.4. For the purpose of photometric mode identification, we adopted $\alpha_{\rm ov}=0.2$ as a reasonable mean value. In our case, however, the choice of $\alpha_{\rm ov}$ has a small effect on the photometric identification of modes.

The theoretical characteristics of pulsations were calculated with the non-adiabatic code of W.\,A.\,Dziembowski \citep{Dzie1977,Pamy1998}. The code includes rotational effects in a perturbative manner, which is justified when the rotational velocity of a modelled star is lower than half of its critical velocity, and pulsational frequencies are much higher than the rotational frequency. Both conditions are fulfilled for all but two modes detected in the program stars (see Sect.\,\ref{star14-16-27}). Theoretical fluxes were calculated using the non-local thermal equilibrium (NLTE) models of \citet{Lanz2007} for solar metallicity, microturbulence velocity $\xi_{\rm t} = 2$ {\kms}, and the non-linear limb-darkening coefficients for NLTE atmospheres provided by \citet{Dasz2011}.

\begin{figure}
\centering
\includegraphics[width=\columnwidth]{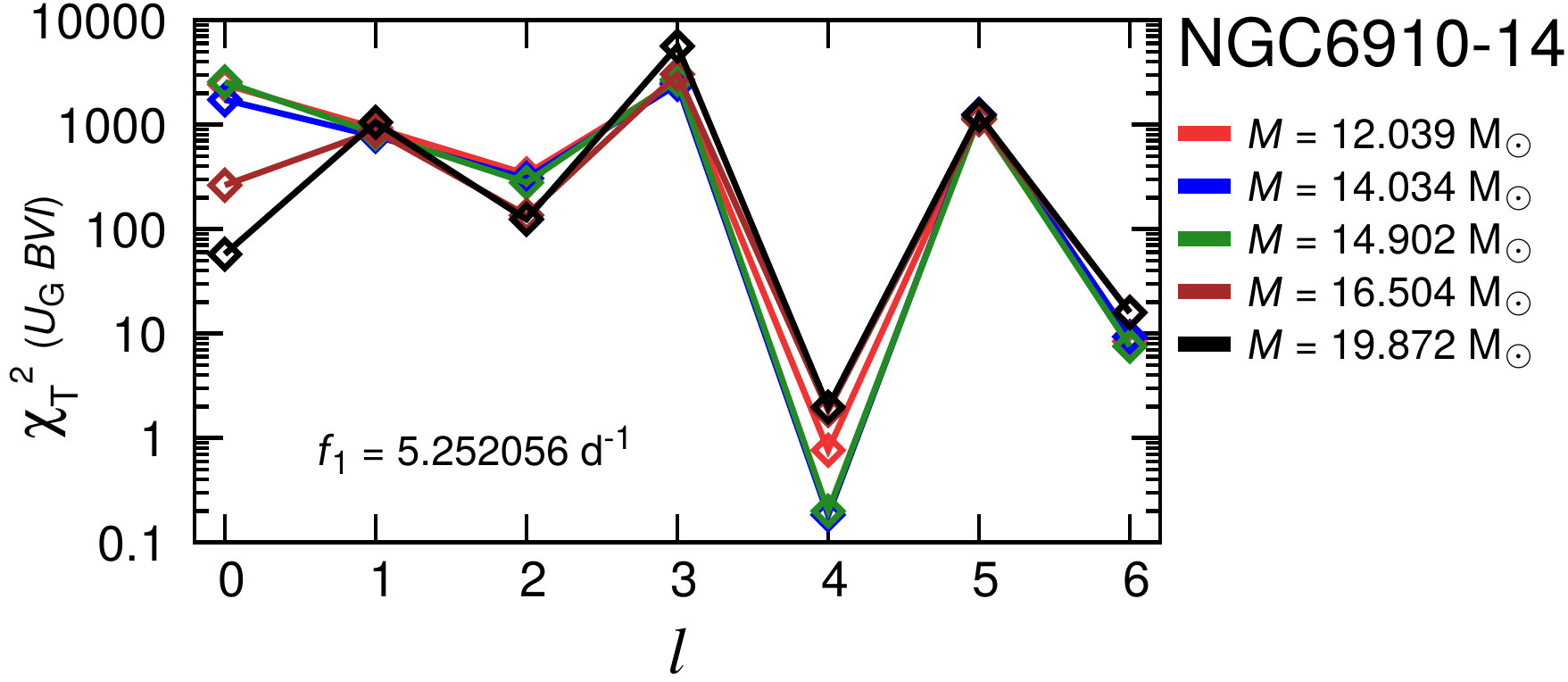}
\caption {$\chi_{\rm T}^2$ vs. $l$ relation for NGC\,6910-14. Passbands in which amplitudes were used for mode identification are given in parentheses. Different colours stand for models with different masses (labelled).}
\label{14_ampident}
\end{figure}

To identify pulsation modes, we used diagnostic diagrams showing the $\chi_{\rm T}^2$ goodness-of-fit parameter defined by \citet{Dasz2009}. For each star, five models were calculated, one located in the centre, and four in the corners of the ($\log T_{\rm eff}$, $\log g$) error box. We considered spherical harmonic degrees up to $l=6$. An example of the $\chi_{\rm T}^2$ versus $l$ relation for the single mode observed in NGC\,6910-14 is given in Fig.\,\ref{14_ampident}. Similar plots for the remaining modes are presented in Fig.\,\ref{16_ampident}\,--\,\ref{36_ampident} in Appendix \ref{Xt_diagrams}. We decided to use only the amplitude ratios to calculate the values of $\chi_{\rm T}^2$ because adding phase differences did not improve the constraints on the identification of $l$. This little diagnostic value of phase differences comes from the fact that for all models considered here, the phase differences are close to zero. In addition, their relative uncertainties are larger than the differences for different models. The results of photometric identification of $l$ are listed in Table \ref{modeidtab}. Multiple identifications indicate that the identification is not conclusive and two or more $l$ result in a similar $\chi_{\rm T}^2$. Depending on the value of $\chi_{\rm T, {\rm min}}^2$, which is the minimum value of $\chi_{\rm T}^2$ of all $l$ and all models considered for a given star, we adopted all identifications that fulfilled the following criteria: If $\chi_{\rm T, {\rm min}}^2<1$ for any model, we adopted all identifications with $\chi_{\rm T}^2 < 2$. If $\chi_{\rm T, {\rm min}}^2>1$, we adopted identifications with $\chi_{\rm T}^2 < 2\chi_{\rm T, {\rm min}}^2$. The values of photometrically identified $l$ in Table \ref{modeidtab} are given in the order of increasing $\chi_{\rm T}^2$. 
\begin{figure}
\centering
\includegraphics[width=\columnwidth]{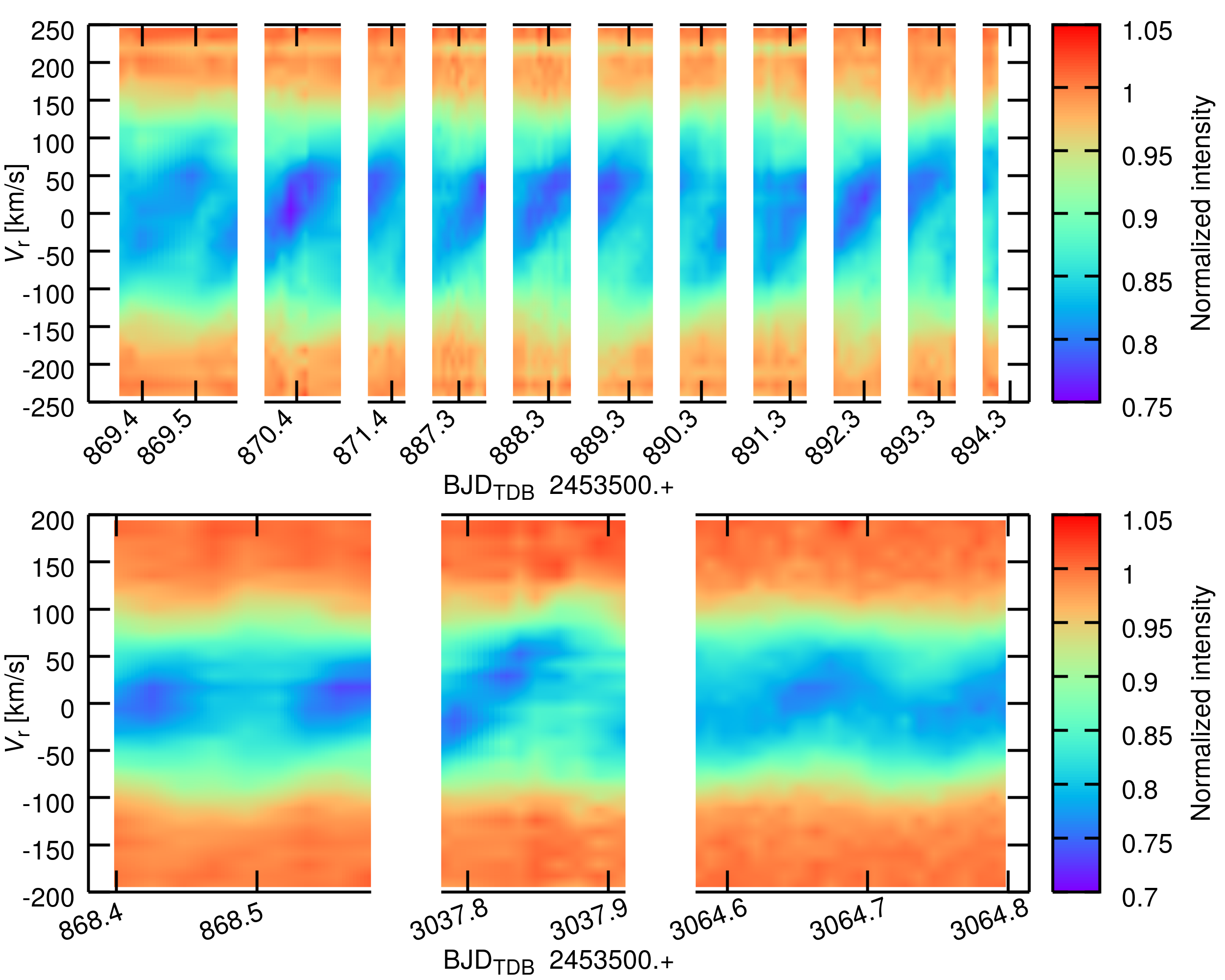}
\caption{Variability of the profile of the \ion{He}{i} 587.56\,nm line in NGC\,6910-14 (top) and NGC\,6910-18 (bottom). For better visibility, the original profiles were smoothed. The time coordinate is not continuous.}
\label{FigProfs}
\end{figure}

Table \ref{modeidtab} shows that $l$ was identified unambiguously for only five modes. In the case of $f_1$ in NGC\,6910-14, identification clearly points to $l=4$ (Fig.\,\ref{14_ampident}). A high value of the spherical harmonic degree like this is unlikely because geometric cancellation reduces the photometric amplitudes of high-$l$ modes \citep{Bal1999,Dasz2002}. Furthermore, this is the only mode detected in NGC\,6910-14, having a relatively large amplitude (8.5~mmag in $V$). Table \ref{atmpartab} shows that the projected equatorial velocity of NGC\,6910-14 is about 150~{\kms}, which is high enough for the pulsation modes to couple. \citet{Dasz2002} showed that at moderate rotation rates (which is the case for stars in NGC\,6910), a strong coupling between modes of $l$ differing by 2 and of the same order $m$ takes place. It is therefore quite possible that the mode in NGC\,6910-14 is in fact the mode with $l<4$ identified as $l=4$ due to rotational coupling. A similar behaviour can be expected for the other modes in the program stars.
\begin{table}
\centering\small
\caption{Summary of the results of photometric and ensemble identification of the spherical harmonic degree $l$ and mode classification for nine program stars in NGC\,6910. Ensemble identifications of $l$, which did not survive second iteration of EnsA (Sect.\,\ref{step-second}), are set in parentheses.}Frequencies marked with a bullet have $V$-filter amplitudes between 1 and 2~mmag, and those marked with an asterisk are below 1~mmag (Table \ref{1stPerPar}).
\label{modeidtab} 
\begin{tabular}{clrccc}
\hline\hline\noalign{\smallskip}
 Star& \multicolumn{2}{c}{Frequency} & Photometric  & Ensemble & Mode\\
 & ID & \multicolumn{1}{c}{[d$^{-1}$]} & $l$ & $l$ &\\
\noalign{\smallskip}\hline\noalign{\smallskip}                
14 & $f_1$ & 5.252056  & 4 & 4,2,3,1 & $p$ or $p/g$\\
\noalign{\smallskip}\hline\noalign{\smallskip}
16 & $f_1$ & 5.202740  & 2 & 2 & $p$\\
 & $f_2$ & 4.174670 & 2 & 2 & $p$\\
 & $f_3$ & 5.846318 & 4,2,6 & 2,1,3,0 & $p$ or $p/g$\\
 & $f_4$ & $\bullet$ 4.588471 & 2 & 1,3 & $p$ or $p/g$\\
 & $f_5$ & $\bullet$ 5.878679 & 6,4 & 2,1,0,3 & $p$ or $p/g$\\
 & $f_6$ & $\ast$ 6.759432 & 4,2,6   & 1,3,2 & $p$\\
 & $f_7$ & $\ast$ 6.314395 & 3,5,0 & 3,0,2,1 & $p$ or $p/g$\\
\noalign{\smallskip}\hline\noalign{\smallskip}
27 & $f_1$ & 6.942973 & 1,2 & 0\,--\,3 & $p$\\
 & $f_2$ & 7.773690 & 6,4 & 0\,--\,3 & $p$ or $p/g$\\
 & $f_3$ & $\bullet$ 1.100920 & 1,2,0 & $>0$ & $g$\\
 & $f_4$ & $\bullet$ 8.433329 & 4,6,2 & 0\,--\,3 & $p$\\
 & $f_5$ & $\bullet$ 7.463826 & 6,4 & 0\,--\,3 & $p$ or $p/g$\\
 & $f_6$ & $\ast$ 7.146300 & 6,4,2 & 0\,--\,3 & $p$\\
 & $f_7$ & $\ast$ 9.262165 & 1,2,0,4,6,5,3 & 0\,--\,3 & $p$\\
\noalign{\smallskip}\hline\noalign{\smallskip}
18 & $f_1$ & 6.154885 & 0 & 0 & $p$ \\
 & $f_2$ & 6.388421 & 2,4 & 3,1,[0] & $p$\\
 & $f_3$ & 6.715615 & 2,4 & 1 & $p$\\
 & $f_4$ & $\bullet$ 5.991368 & 6,4,2 & 2,0 & $p$ or $p/g$\\
 & $f_5$ & $\bullet$ 8.871393 & 1,0,2 & 1,3,2 & $p$\\
 & $f_6$ & $\bullet$ 7.074318 & 0,5 & 2,0,[3] & $p$ or $p/g$\\
 & $f_7$ & $\bullet$ 6.658287 & 6,4 & 1,[3] & $p$\\
 & $f_8$ & $\bullet$ 7.879447 & 2,1 & 0,1,2 & $p$\\
 & $f_9$ & $\ast$ 7.755512 & 6,4 & 0,1,3,2 &$p$ or $p/g$\\
 & $f_{10}$ & $\ast$ 7.219347 & 4,6 & 2,0,3 & $p$\\
 & $f_{11}$ & $\ast$ 5.612781 & 1,2,0,4  & --- & ---\\
 & $f_{12}$ & $\ast$ 6.579093 & 0,5,3,1,2  & 1,3 & $p$\\
\noalign{\smallskip}\hline\noalign{\smallskip}
25 & $f_1$ & $\bullet$ 7.141122 & 4,2,6,5 & 2,3 & $p/g$\\
 & $f_2$ & $\ast$ 8.120426 & 0,1,2,5,4,3,6 & --- & ---\\
\noalign{\smallskip}\hline\noalign{\smallskip}
41 & $f_1$ & 9.845570 & 0,1 & 2,0,3 & $p$\\
\noalign{\smallskip}\hline\noalign{\smallskip}
34 & $f_1$ & 3.826208 & 4,6 & 2,3 & $g$\\
 & $f_2$ & $\ast$ 11.548531 & 2,4,1,6,0 & 1,3 & $p$\\
\noalign{\smallskip}\hline\noalign{\smallskip}
38 & $f_1$ & 4.793720 & 4,6,2 & 2 & $g$\\
 & $f_2$ & $\ast$ 12.909208 & 0,1,2,5,3,4,6 & 2,0,3 & $p$\\
\noalign{\smallskip}\hline\noalign{\smallskip}
36 & $f_1$ & 0.363151 & 1,0,2  & $>0$ & $g$ \\
 & $f_2$ & $\bullet$ 5.185733 & 4,6,2 & 2 & $g$ \\
 & $f_3$ & $\bullet$ 4.725566 & 2,4,6,1 & 3& $g$ \\
\noalign{\smallskip}\hline                  
\end{tabular}
\end{table}

\subsection{Spectroscopic identification of modes}\label{SpeId_ch}
We were only able to apply spectroscopic mode identification for two program stars, NGC\,6910-14 and 18, which had sufficiently large number of spectra gathered (Sect.~\ref{SpeObs_ch}). Because the variability of line profiles through pulsations is clearly visible for both stars (Fig.\,\ref{FigProfs}), we used the Fourier parameter fit (FPF) method developed by \citet{Zima2006a}. The FPF method takes advantage of the amplitudes and phases derived from the time series of intensities calculated along the profile of a spectral line. These amplitudes and phases are then compared with the synthetic ones, calculated from the models. The method is similar to the pixel-by-pixel method developed by \cite{Mante2000}, but the FPF allows estimating the statistical quality of the fits. If the spectra have S$/$N $\gtrsim$ 200, some stellar parameters, for example the inclination of the rotational axis, can be constrained in addition to $l$ and $m$. Unfortunately, our spectra have S$/$N in the range between 50 and 100, which is the low limit for the application of the FPF method \citep{Zima2006a}. The advantage of the FPF method is that good constraints on $m$ (even better than on $l$) can sometimes be obtained. The FPF method was successfully applied for mode identification in many stars  \citep{Zima2006b,Des2009,Bri2012}. Becauseit is implemented in the frequency analysis and mode identification for asteroseismology (FAMIAS\footnote{Mode
identification results obtained with the software package FAMIAS developed
in the framework of the FP6 European Coordination Action HELAS (http://www.helas-eu.org/)}
) package \citep{Zima2008}, we used this package in the present work. 
\begin{figure*}
\centering
\includegraphics[width=\textwidth]{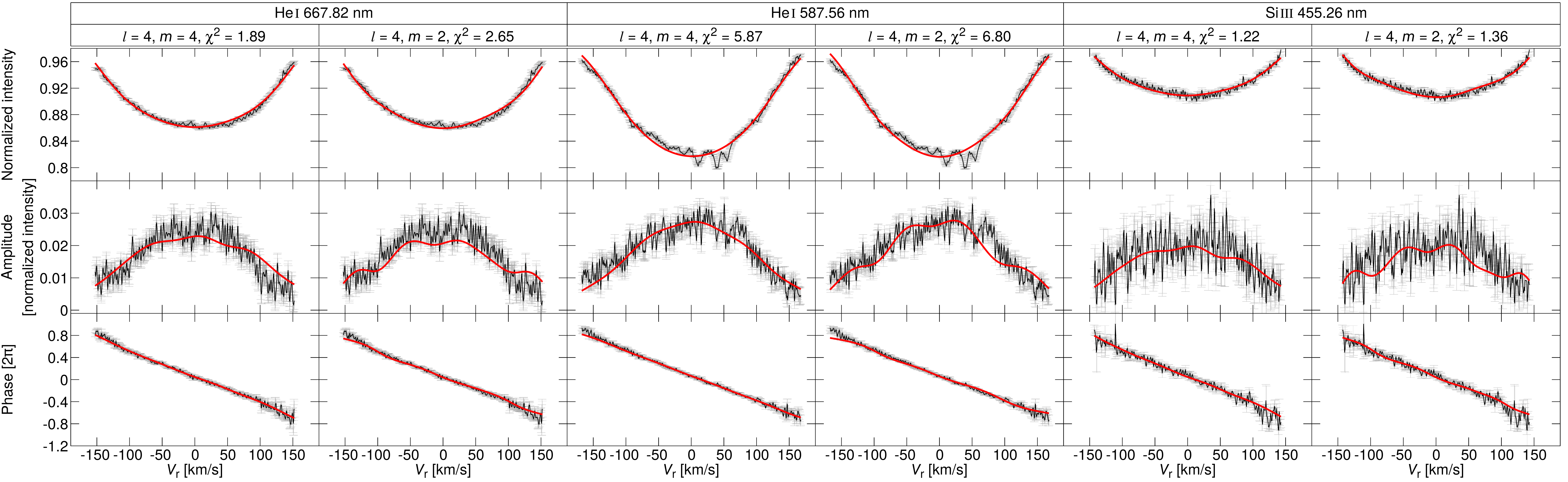}
\caption{Results of the FPF fits of $f_1$ to \ion{He}{i} 667.82\,nm, \ion{He}{i} 587.56\,nm, and \ion{Si}{iii} 455.26\,nm spectral lines of NGC\,6910-14. The top panels show the line profiles, the middle panels the amplitudes, and the bottom panels the phases. Observed profiles, amplitudes, and phases are plotted with black lines, their errors are marked with grey bars, and model fits are plotted with red lines. The profile of the \ion{He}{i} 587.56\,nm line near the core is disturbed by telluric lines.}
\label{FigFPF14}
\end{figure*}

Because our spectra of NGC\,6910-14 and 18 had moderate S$/$N, we used only three strong lines, \ion{He}{i} 587.56\,nm, \ion{He}{i} 667.82\,nm, and \ion{Si}{iii} 455.26\,nm to identify modes. The last line is considerably weaker than the first two. In modelling with FAMIAS, the surfaces of stars were divided into 10\,000 segments, which is sufficiently large to identify high-$l$ modes in the presence of fast rotation \citep{Zima2008}. The ranges of masses and radii for the models (calculated with the Warsaw-New Jersey code) of the two stars were assumed to satisfy the adopted error boxes of $T_{\rm eff}$ and $\log g$ (Table \ref{atmpartab}). The $V_{\rm eq}\sin i$ parameters were fixed on values taken from the same table.

\begin{figure}[!t]
\centering
\includegraphics[width=\columnwidth]{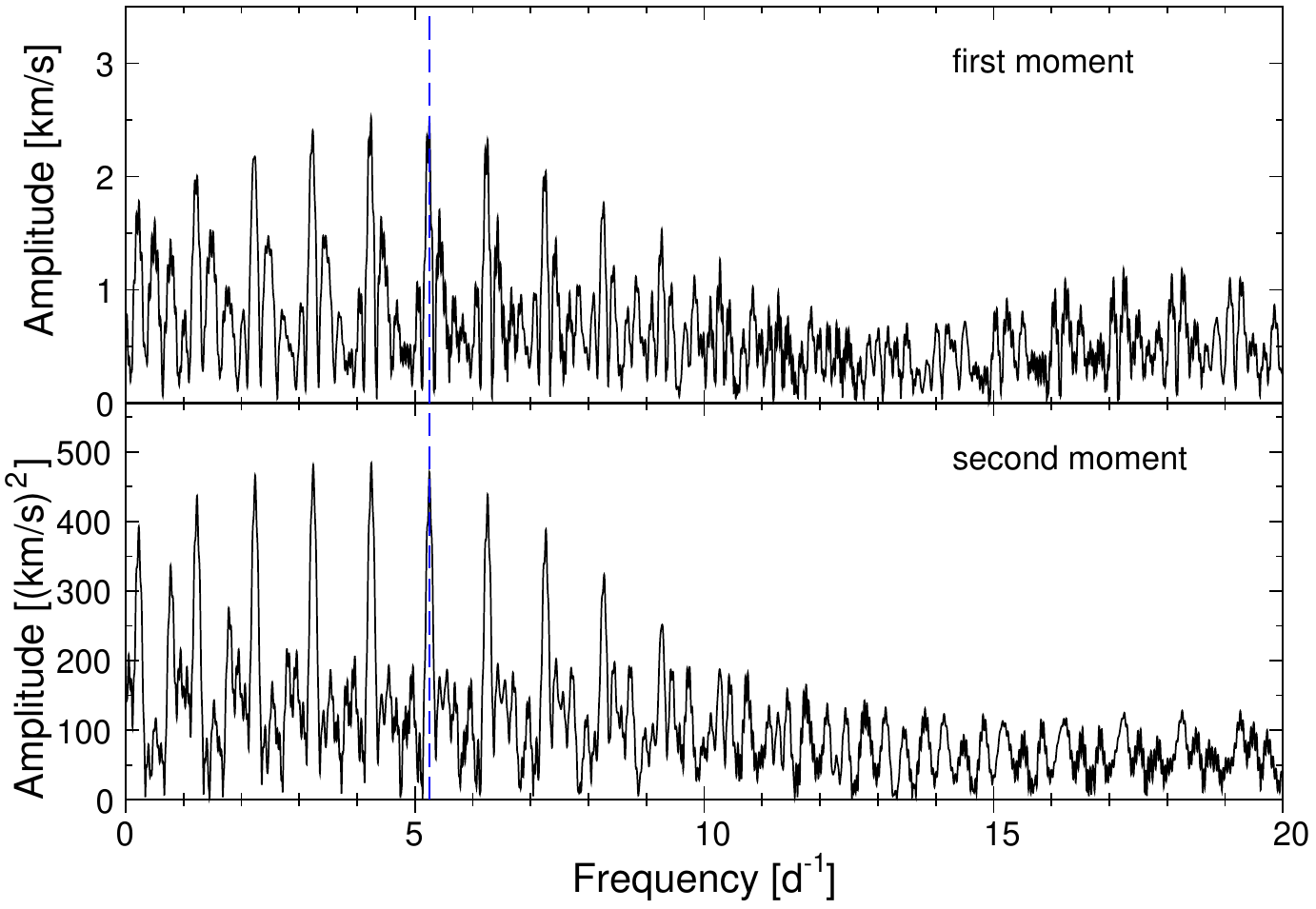}
\caption{Frequency spectra of the average first (top) and second (bottom) moments of two helium line profiles in the spectra of NGC\,6910-14. The vertical dashed line marks the frequency $f_1=5.252$~d$^{-1}$ detected in photometry.}
\label{14-mom}
\end{figure}

The variability of the line profiles for NGC\,6910-14 is shown in Figs.\,\ref{FigProfs} and \ref{FigFPF14}. The variability is also well pronounced in the analysis of the profile moments, in which the $n$th normalised moment, $\langle v^n\rangle$, is defined as \citep{Bri2003}
\begin{equation}\label{moment}
\langle v^n\rangle(t) = \frac{\int_{v_{\rm min}}^{v_{\rm max}}v^n I(v,t)\mbox{ d}v}{\int_{v_{\rm min}}^{v_{\rm max}}I(v,t)\mbox{ d}v},
\end{equation}
where $v$ is the observed velocity of a point on the stellar surface, and $I(v,t)$ is the intensity of a line profile. The integral is calculated over the full line profile, between $v_{\rm min}$ and $v_{\rm max}$. The denominator in Eq.\,(\ref{moment}) corresponds to the equivalent width of a line. The first moment, $\langle v^1\rangle$, is a good measure of the radial velocity. Despite strong daily aliasing, periodograms of the first and second moments of the helium lines (Fig.\,\ref{14-mom}) clearly show variability with the frequency detected in photometry, $f_1=5.252056$~d$^{-1}$. Radial velocities calculated as average $\langle v^1\rangle$ from the \ion{He}{i} 587.56 and 667.82~nm lines are listed in Table \ref{RVs-14}. The amplitude of radial velocities obtained from the sine-curve fit of $f_1$ is 2.46\,$\pm$\,0.20\,{\kms}. 
The frequency $f_1$ was also detected from the variability of intensities along the line profiles. Similarly to the analysis of moments, no other variability than with $f_1$ was found. We searched for the best-fit model with FAMIAS allowing $l$ in the range between 0 and 4. Two best-matched models were obtained for $(l,m)=(4,+4)$ and $(l,m)=(4,+2)$ (Fig.\,\ref{FigFPF14}). Although $\chi^2$ is slightly lower for the former model, the latter seems to reproduce the shape of amplitude along profiles of spectral lines better (middle panels of Fig.\,\ref{FigFPF14}).

\begin{figure}
\centering
\includegraphics[width=\columnwidth]{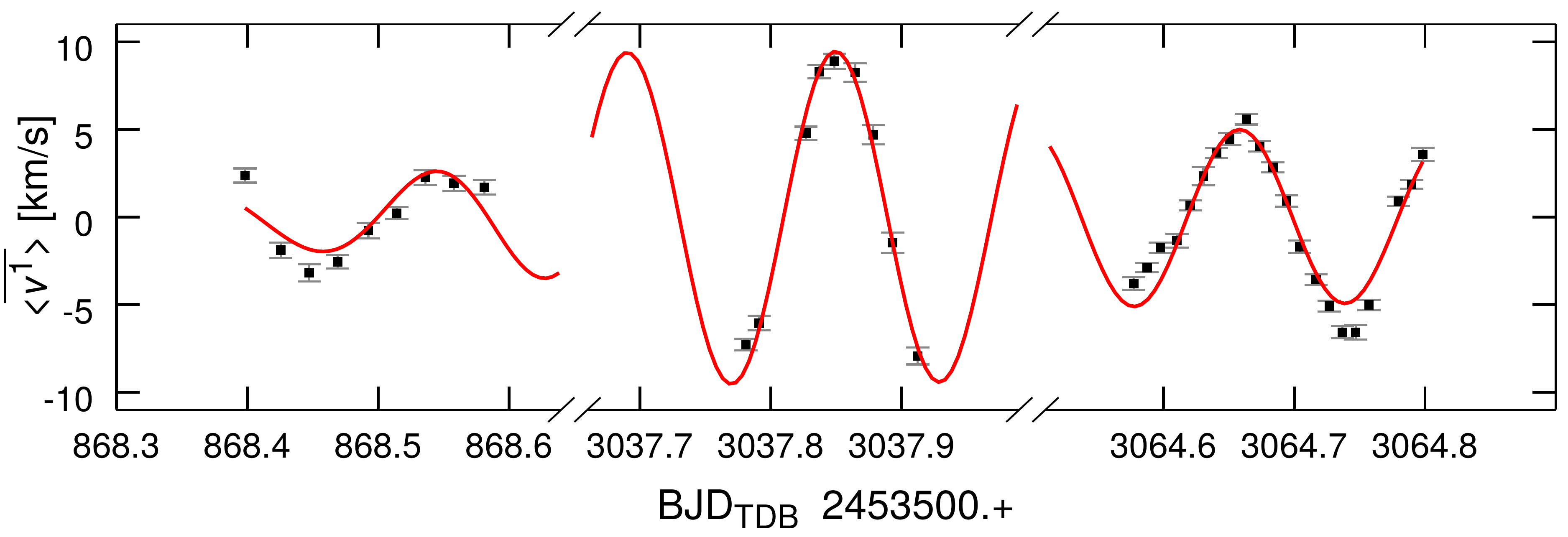}
\caption{Radial velocity curve of NGC\,6910-18 (average $\langle v^1\rangle$ from two \ion{He}{i} lines) compared with the model including $f_1$, $f_2$ , and $f_3$ (red line). The observations accounted for the 14.3\,{\kms} difference between the 2007 and 2013 radial velocities. The time coordinate is not continuous.}
\label{FigW18_1mom}%
\end{figure}

\begin{figure}
\centering
\includegraphics[width=\columnwidth]{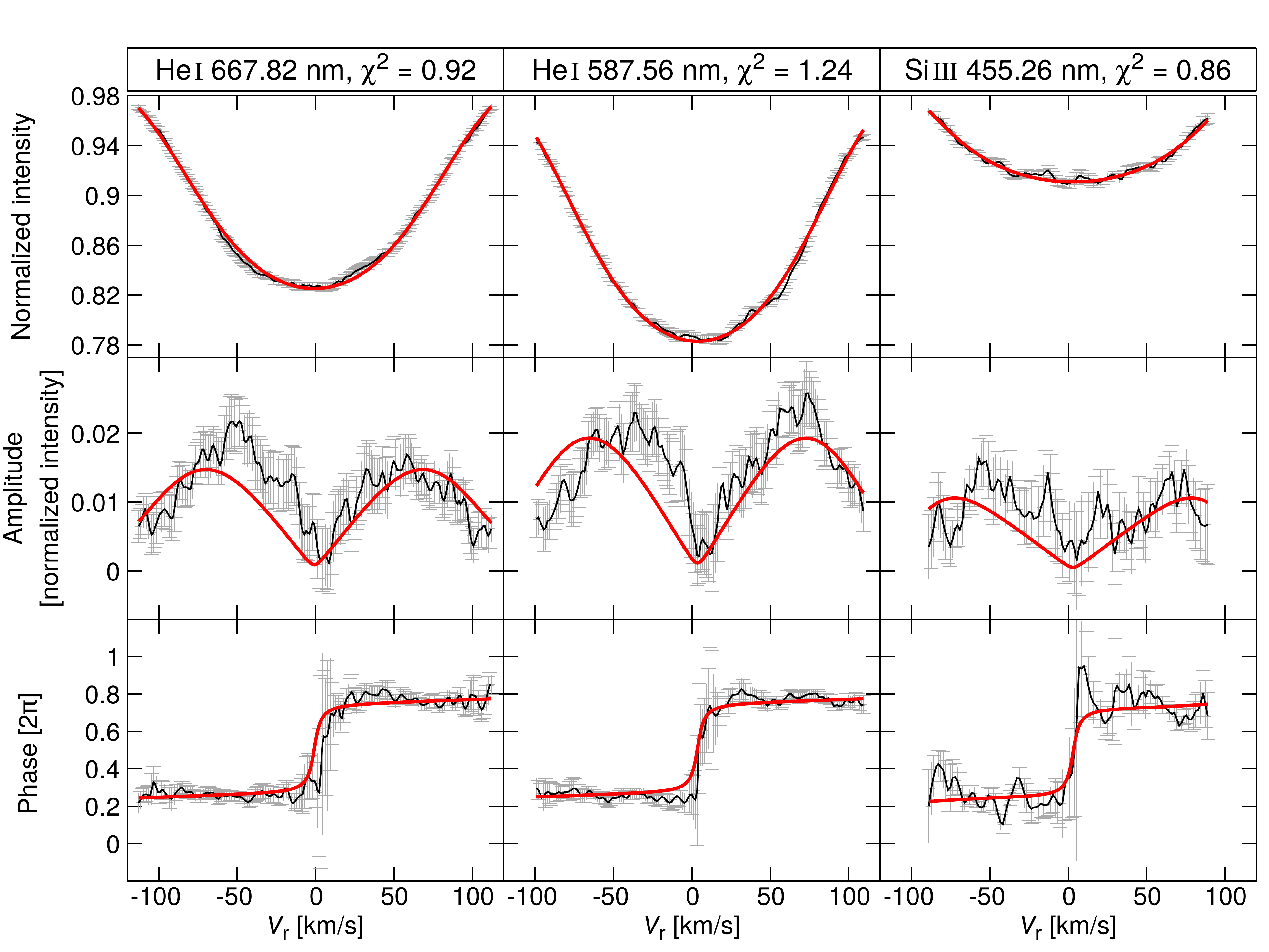}
\caption{Same as in Fig.\,\ref{FigFPF14}, but for the $f_1$ (radial) mode in NGC\,6910-18.}
\label{FigFPF18}
\end{figure}

Similarly to NGC\,6910-14, the spectral lines of NGC\,6910-18 show prominent variability along the line profiles (Fig.\,\ref{FigProfs}). To obtain less noisy radial velocities, we averaged the first moments for the two helium lines, He\,{\sc i} 587.56\,nm  and He\,{\sc i} 667.82\,nm. We did not use the results obtained for the \ion{Si}{iii} 455.26\,nm line because they showed much larger scatter. For reference, however, the results for the \ion{Si}{iii} line are shown in Figs.\,\ref{FigFPF14} and \ref{FigFPF18}. The radial velocities calculated from the first moments are shown in Fig.\,\ref{FigW18_1mom} and are also provided in Table \ref{RVs-18}. The spectroscopic data of NGC\,6910-18 are very sparse, but their distribution in time allows resolving the three strongest modes, $f_1$, $f_2$ , and $f_3$, assuming periods obtained from the analysis of the photometry (Table \ref{1stPerPar}). The resulting amplitudes of the radial velocity variations related to these three modes are equal to $A_1=6.23\,\pm\,0.30$\,{\kms}, $A_2=3.22\,\pm\,0.28$\,{\kms} , and $A_3=1.56\,\pm\,0.25$\,{\kms}. 

For NGC\,6910-18, we used the FPF method in the same way as for NGC\,6910-14. At the beginning, our model included the three strongest modes, $f_1$, $f_2$ , and $f_3$ (Table \ref{1stPerPar}). Unfortunately, the obtained solutions incorrectly reproduced line profiles, amplitudes, and phases. We therefore assumed a simpler model, consisting of only a single dominant mode with frequency $f_1=6.154885$\,d$^{-1}$. Although the observed dependencies (Fig.\,\ref{FigFPF18}) are not perfectly reproduced by the model because other modes are present, the $l=0$ identification for $f_1$ is confirmed without any doubt: the amplitude drop in the centre of the profile and the phase switch by $\pi$ are the key attributes of a radial mode (e.g.~Fig.~1 in \citealt{Zima2006a} or Fig.~3 in \citealt{Bri2012}).

Summarising the results of the photometric and spectroscopic identifications, we see that only five modes were identified unambiguously, including the radial mode in NGC\,6910-18 identified as such by both photometric and spectroscopic methods. Interestingly, the single mode of relatively high amplitude observed in NGC\,6910-14 was consistently identified as $l=4$ both from photometry and spectroscopy.

\subsection{Crucial constraint: cluster age}\label{step-age}
One of the most important foundations of the EnsA is the assumption of the coevality of cluster members. Even if the number of identified modes is small at the beginning, as in the present case, the modes may place a tight constraint on the cluster age. Potentially, this can start an iterative process of narrowing down the cluster age using pulsational models of all stars. A by-product of this procedure can also be the exclusion of some multiple identifications, which may eventually lead to the additional unambiguous identification of more modes.

To calculate evolutionary and pulsation models, we used the same codes as described in Sect. \ref{PhoId_ch} assuming $Z=Z_\odot$. Because the $\log g$ values for program stars determined in Sect.\,\ref{AtmPar_ch} are subject of relatively large uncertainties, we decided to search for models using only the limits of much better determined $T_{\rm eff}$. The models were calculated for five values of $\alpha_{\rm ov}$, between 0.0 and 0.4, with a step of 0.1. Next, we adopted $V_{\rm eq}$ equal to mean $\langle V_{\rm eq}\rangle=$ 170\,{\kms} (Sect.\,\ref{SpeId_ch}) for all stars except for NGC\,6910-14 and 18. For NGC\,6910-14 and 18 we calculated models for two limiting values of $V_{\rm eq}$: one equal to $V_{\rm eq}\sin i$ (Table \ref{atmpartab}), the other to half the critical velocity, an upper limit for the rotation effects included in our pulsational models (Sect.\,\ref{PhoId_ch}). The two values were equal to 97\,{\kms} and 250\,{\kms} for NGC\,6910-18, and 170 and 220\,{\kms} for NGC\,6910-14. The final grid included 198\,793 pulsation models and 499\,083 evolutionary models. The masses for these models ranged between 4.4 and 21 M$_\odot$. The models that reproduced the observed frequencies were interpolated using $T_{\rm eff}$ as a free parameter.
\begin{figure}[!ht]
\centering
\includegraphics[width=\columnwidth]{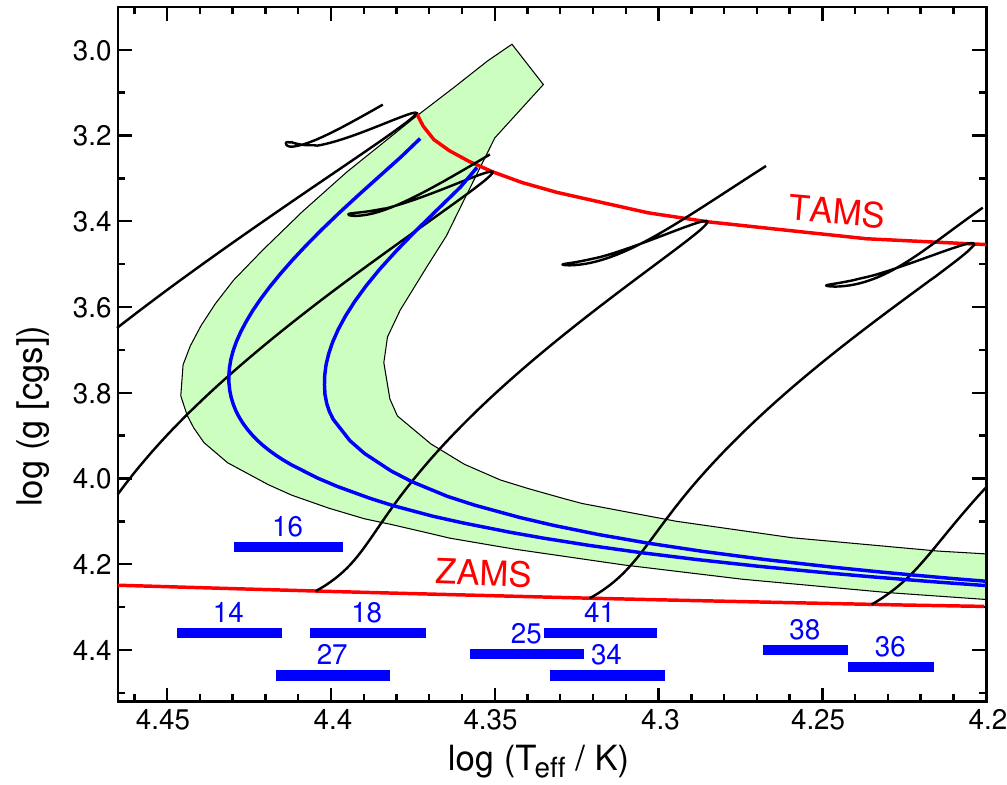}
\caption{Kiel diagram for models of program stars in NGC\,6910. The green area limits the location of models with ages between 9.25 and 11.61~Myr in the core hydrogen-burning phase (main sequence) with $0.0<\alpha_{\rm ov}<0.4$ and $97\,\mbox{km\,s}^{-1}< V\sin i<250$\,{\kms}. The red lines stand for the ZAMS and TAMS phase. Black lines are evolutionary tracks for 5, 7, 10, 15, and 20~$M_\odot$. Blue lines are isochrones for 9.25 and 11.61~Myr. Evolutionary tracks and isochrones are shown for $V_{\rm eq}=$~170\,{\kms} and $\alpha_{\rm ov}=$~0.2. For each program star (labelled with number) the adopted range of $T_{\rm eff}$ (Table \ref{atmpartab}) is marked by a blue bar in the lower part of the plot.}
\label{Kiel0}
\end{figure}

In seismic modelling, the most valuable are radial modes because their frequencies are much less sensitive to rotational velocity than non-radial modes. Fortunately, one radial mode has been unambiguously identified in NGC\,6910-18. We therefore started EnsA with NGC\,6910-18, as this is the only program star with an unambiguously identified radial mode ($f_1$). In the range of adopted $T_{\rm eff}$ we found that models that reproduce $f_1$ as a radial mode have ages in the range between 9.25 and 16.75~Myr. The next constraint for the age of NGC\,6910 came from NGC\,6910-14, the brightest and the most massive of the program stars. From the evolutionary models falling in the adopted range of $T_{\rm eff}$ for this star, we could conclude that it cannot be older than 11.61~Myr. This narrowed the range of the cluster age to 9.25\,--\,11.61~Myr. We adopted this age range as the main constraint in the discussion of mode identification for all program stars presented in the next subsection. The considered models are presented in the $\log g$ versus $\log T_{\rm eff}$ (called `Kiel') diagram (Fig.\,\ref{Kiel0}).

\subsection{Modelling individual stars}\label{step-mid}
After constraining the age of the cluster (Sect.\,\ref{step-age}), we proceeded to exclude some mode identifications and narrow stellar parameters for program stars using evolutionary and pulsational models. For a given star, we sought models that simultaneously satisfied the following five criteria: (i) the age of the model was between 9.25 and 11.61~Myr, (ii) $T_{\rm eff}$ of the model was in the range defined in Table \ref{atmpartab}, (iii) the model was in the core hydrogen-burning phase (main sequence), (iv) the theoretical frequency of a given mode was the same as observed, and (v) the instability parameter $\eta$ \citep{Stelli1978,Dzie2008} of the mode was larger than $-0.5$. The criteria (i) to (iii) define a subset of evolutionary models for a given star. In Fig.\,\ref{Kiel0} these models are located in the common part of the green area and the strip defined by $T_{\rm eff}\pm \Delta T_{\rm eff}$ (Table \ref{atmpartab}).
\begin{figure*}[!ht]
\centering
\includegraphics[width=\textwidth]{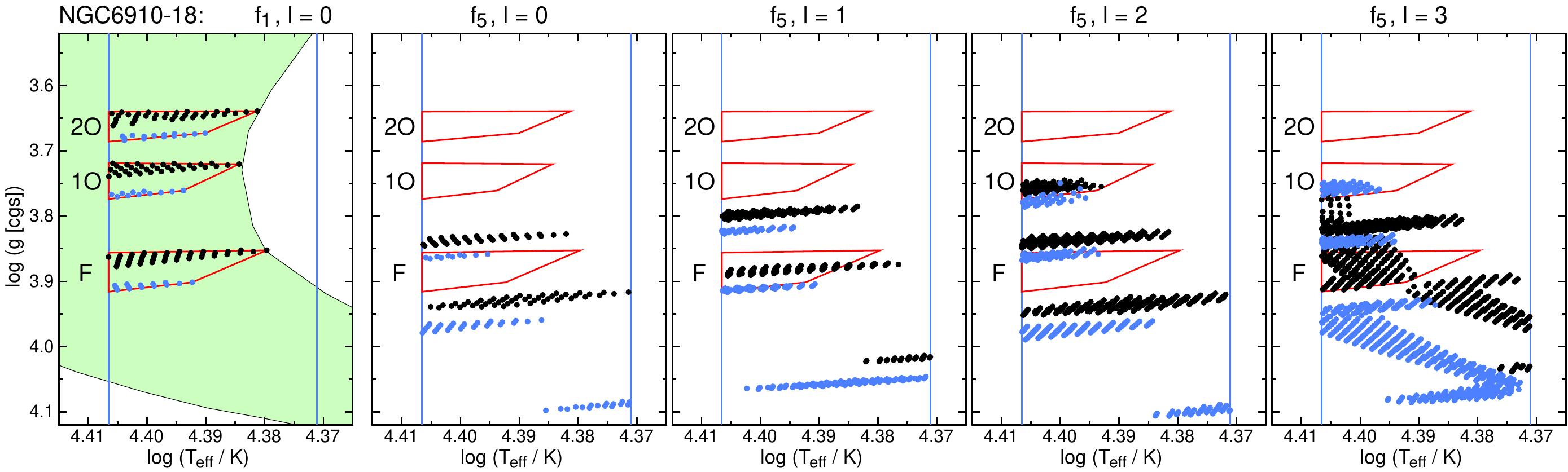}
\caption{Kiel diagrams for models of NGC\,6910-18. The leftmost panel shows models that satisfy all criteria, (i) to (v), for the radial mode ($f_1$). Black and blue dots correspond to modes with rotational velocities of 250 and 97~{\kms}, respectively. The green area is the same as in Fig.\,\ref{Kiel0}; blue vertical lines correspond to the adopted $T_{\rm eff}$ limits for NGC\,6910-18. The models define three areas (shown with red boundaries) corresponding to $f_1$ identified as fundamental (F), first overtone (1O), and second overtone (2O) radial mode. The remaining four panels show the location of models satisfying all criteria for the $f_5$ mode assuming (from left to right) $l=0$, 1, 2, and 3. The colours of the dots have the same meaning as in the leftmost panel.}
\label{Kiel-18}
\end{figure*}

Criteria (iv) and (v) require calculation of pulsational models. For modes with unambiguously identified $l$, we can directly compare observed and theoretical frequencies, considering for radial modes all possible radial orders $n$, for non-radial modes, all possible $n$ and $m$. For modes that have no unambiguous mode identification, all values of $l$ (and all allowable $n$ and $m$) need to be considered. The situation is complicated by the possible presence of rotational coupling. As shown by \cite{Dasz2002} in their Figs.~4 and 6, in the presence of rotational coupling, an $l=2$ mode can be identified as a mode with any $l$ between 0 and 4 if the amplitude ratio is used to constrain $l$. This possibility makes the results of our photometric mode identification (Table \ref{modeidtab}) less certain. We therefore decided to proceed in the following way. Unless explicitly indicated, modes with $l\leqslant 3$ were considered, regardless of the photometric identification. In addition, when defining the overlapping areas of models for different modes in the Kiel diagram, we used only modes with amplitudes exceeding 2~mmag in $V$. The only exceptions were stars that do not have modes like this (NGC\,6910-25 and 36)\footnote{NGC\,6910-36 has a mode with a $V$ amplitude exceeding 2~mmag in $V$, but with a frequency of 0.363~d$^{-1}$. This means it is a $g$ mode, which gives no constraint on mode identification.}. The choice of $l\leqslant 3$ can be justified by the averaging effect for modes with higher $l$ \citep{Dasz2002}.

For criterion (v) we required $\eta>-0.5$ instead of $\eta>0$. This needs an explanation. In general, $\eta>0$ means that a mode is unstable, and $\eta<0$ signifies that a mode is stable. The reason is that for real stars the current opacities do not always explain the excitation of the observed modes. The adoption of (v) was therefore a compromise between the real situation in seismic modelling of $\beta$~Cep stars and formal interpretation of $\eta$. We discuss the stability of modes in program stars in Sect.\,\ref{stability}.
For stars with multiple modes, additional constraints resulted from the requirement that models for modes satisfying the criteria have to overlap in the Kiel diagram.

\subsubsection{NGC\,6910-18}\label{star18}
This star is crucial for the EnsA in NGC\,6910 because its radial mode ($f_1$) is unambiguously identified. The leftmost panel of Fig.\,\ref{Kiel-18} shows the result of applying all criteria for the radial mode assuming two extreme values of $V_{\rm eq}$, 97 and 250~{\kms}. This allows defining three narrow areas in the Kiel diagram corresponding to the fundamental (F) and two lowest overtone (1O and 2O) pulsations. (Higher radial overtones are excluded by criterion (v) because they are severely damped.) For the remaining 11 modes detected in this star (none of them is identified unambiguously), we considered, as explained above, all models with $l=0$, 1, 2, and 3, and  plotted them in the Kiel diagram. We then verified which models fall in the areas defined by the radial mode.
\begin{figure*}[!ht]
\centering
\includegraphics[width=0.9\textwidth]{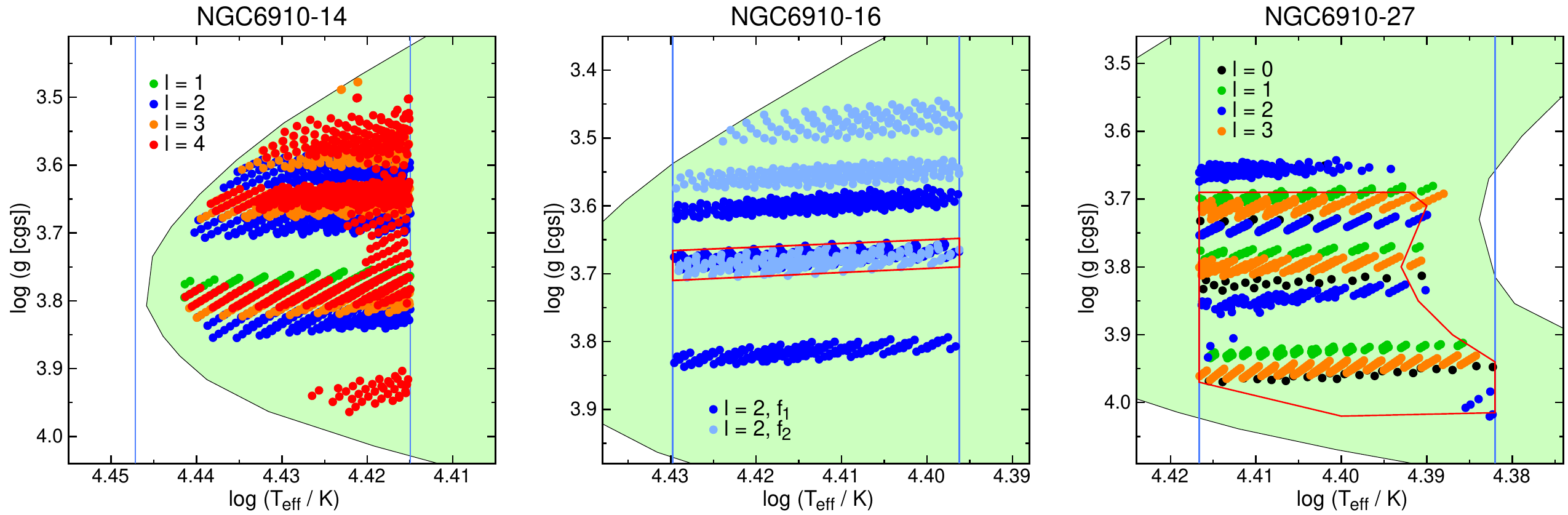}
\caption{Kiel diagrams for models of NGC\,6910-14 (left), 16 (middle), and 27 (right). The green area is the same as in Fig.\,\ref{Kiel0}. Blue vertical lines correspond to the adopted $T_{\rm eff}$ limits from Table \ref{atmpartab}. For these program stars, pulsational models for only a single rotational velocity $V_{\rm eq}=170$~{\kms} were considered. Symbols with different colours in left and right panels correspond to different $l$ (labelled). For NGC\,6910-16 (middle panel), dots represent models for which two observed frequencies, $f_1$ (dark blue) and $f_2$ (light blue), are reproduced as $l=2$ modes. The red line delimits the area common to models that reproduce both these modes. For NGC\,6910-27 (right panel), dots represent models for which the frequency $f_1$ is reproduced. For clarity, models reproducing $f_2$ are not shown. Instead, the area common to models reproducing both $f_1$ and $f_2$ is enclosed with a red line.}
\label{Kiel-14-16-27}
\end{figure*}

As an example, we show in the four panels of Fig.\,\ref{Kiel-18} the results of this procedure for $f_5= 8.871393$~d$^{-1}$. The only models that reproduce $f_5$ as a radial ($l=0$) mode fall in the area defined by $f_1$ as an F mode. However, these are $V_{\rm eq}=97$\,{\kms} models for $f_5$ , and they are located in the upper part of the F area that contains models with $V_{\rm eq}=250$\,{\kms} for $f_1$. This discrepancy allows us to exclude $l=0$ for $f_5$. The next panels show that the possible identifications for $f_5$ are $l=1$ ($f_1$ is F), $l=2$ ($f_1$ is 1O), and $l=3$ ($f_1$ is either F or 1O). Summarising, only $l=1$, 2, and 3 are possible for $f_5$. We call this identification of $l$ `ensemble $l$'.

An important result of the comparison shown in Fig.\,\ref{Kiel-18} is that the 2O for $f_1$ can be excluded, as none of the models for $f_5$ falls into the area defined for 2O. A similar comparison for $f_8$ also excludes 2O for $f_1$. We could therefore perform another iteration of our procedure excluding all identifications in which the agreement between a given mode and $f_1$ was obtained for the 2O-mode area. The final result of this ensemble identification is given in the fifth column of Table \ref{modeidtab}. For multiple identifications, the order of ensemble $l$ corresponds to the size of the overlapping areas. For $f_3$ we obtain $l=1$. On the other hand, no mode with $l\leqslant3$ could be agreed for $f_{11}$. Because this mode has an amplitude below 1~mmag, a possible explanation is that its frequency is wrong by a daily alias. The seismic models also allow us to classify the modes based on the ratio of the kinetic energy in the $g$-mode propagation zone to the total kinetic energy of the mode. This classification is given in the last column of Table \ref{modeidtab}. The designation $p/g$ stands for a mixed mode, that is,~a mode that has a $g$-mode character in the stellar interior and a $p$-mode character in the envelope. These modes are common in $\beta$~Cep stars that are advanced in their main-sequence evolution.

The presented procedure is not aimed at finding a seismic model that would reproduce all observed frequencies. Such a detailed seismic modelling is beyond the scope of this paper, although the results of the ensemble mode identification can be used as a starting point for this type of modelling.

\subsubsection{NGC\,6910-14, 16, and 27}\label{star14-16-27}
The next three stars, NGC\,6910-14, 16, and 27 are among the hottest program stars ($T_{\rm eff}> 23\,000$~K) and the evolutionary models that needed to be considered for them cover a relatively wide range of $\log g$ in the Kiel diagram (Fig.\,\ref{Kiel0}). Their variability (and that of NGC\,6910-18) has been discovered by \cite{Zibi2004}.

NGC\,6910-14 is the brightest and most massive program star with a single mode identified as $l=4$, both from photometry (Table \ref{modeidtab}) and spectroscopy (Sect.\,\ref{SpeId_ch}). Because of the large photometric amplitude and the identification as $l=4$, the mode is, as discussed in Sect.\,\ref{SpeId_ch}, a good candidate for a rotationally coupled mode. Therefore, we decided to plot in the Kiel diagram for this star (Fig.~\ref{Kiel-14-16-27}) not only modes with $l=4$, but also modes with $l\leqslant3$. We found that it is not possible to reproduce the frequency $f_1$ by the allowable models when a radial mode is assumed. For $l = 1$, 2, 3, and 4, such models were found. Because only one mode is observed in NGC\,6910-14, no additional constraints as to the mode identification could be obtained.

Seven modes were discovered in NGC\,6910-16, including three that are identified unambiguously from photometry as $l=2$ modes: $f_1$, $f_2$, and $f_4$. Because $f_4$ has a $V$-filter amplitude below 2~mmag and therefore its photometric identification is less certain, we used only $f_1$ and $f_2$ when we searched for the region in the Kiel diagram in which models reproducing both frequencies overlap. Figure \ref{Kiel-14-16-27} shows that there is only a relatively narrow overlapping region (marked with red borders) in which models satisfy the criteria for both modes. We can therefore proceed in the same way as for NGC\,6910-18 and search for limitations for all the other modes that are excited in this star. The resulting ensemble identification of $l$ is given in Table \ref{modeidtab}. Because the nearly horizontal sequences in the middle panel of Fig.\,\ref{Kiel-14-16-27} correspond to different radial orders $n$, we can identify them as $n=0$ for $f_1$ and $n=-1$ for $f_2$.

None of the seven modes detected in NGC\,6910-27 is identified unambiguously by means of the photometric method. Only the two strongest modes, $f_1$ and $f_2$, have amplitudes exceeding 2~mmag. We therefore plotted all models that reproduce the two strongest modes, $f_1$ and $f_2$, allowing for $l\leqslant3$. This defines a relatively large area in the Kiel diagram (right panel in Fig.\,\ref{Kiel-14-16-27}) that is covered by the allowable models. We derived the ensemble $l$ for the other six modes by comparing the location of all models that reproduce a given frequency with this area. Table \ref{modeidtab} shows that no constraint for $l$ has been obtained for modes that have been classified as $p\text{}$ or mixed $p/g\text{}$ modes.

We made no attempt to reproduce $f_3$ with our models; its frequency equal to 1.100920~d$^{-1}$ clearly indicates it is a $g\text{}$ mode. As explained in Sect.\,\ref{PhoId_ch}, the pulsational code we used included rotational effects in a perturbative manner, therefore the modelling was appropriate only if pulsational frequencies were much higher than the rotational frequency. This condition is not fulfilled for $f_3$ in NGC\,6910-27 and $f_1$ in NGC\,6910-36. Consequently, we did not attempt to identify $l$ for these two apparent $g$ modes. 

\subsubsection{Remaining five program stars}
The remaining five program stars (NGC\,6910-25, 34, 36, 38, and 41) were discovered as variable during the 2005\,--\,2007 campaign. All but NGC\,6910-41 are multiple pulsators. They are cooler (16\,000~K $<T_{\rm eff}<$ 23\,000~K) than the stars discussed in Sects.~\ref{star18} and \ref{star14-16-27}. In consequence, the area of the models that satisfies all criteria for these stars is relatively narrow and almost horizontal in the Kiel diagram. One of the consequences of this fact is a good constraint on their surface gravities.

\begin{figure}[!t]
\centering
\includegraphics[width=\columnwidth]{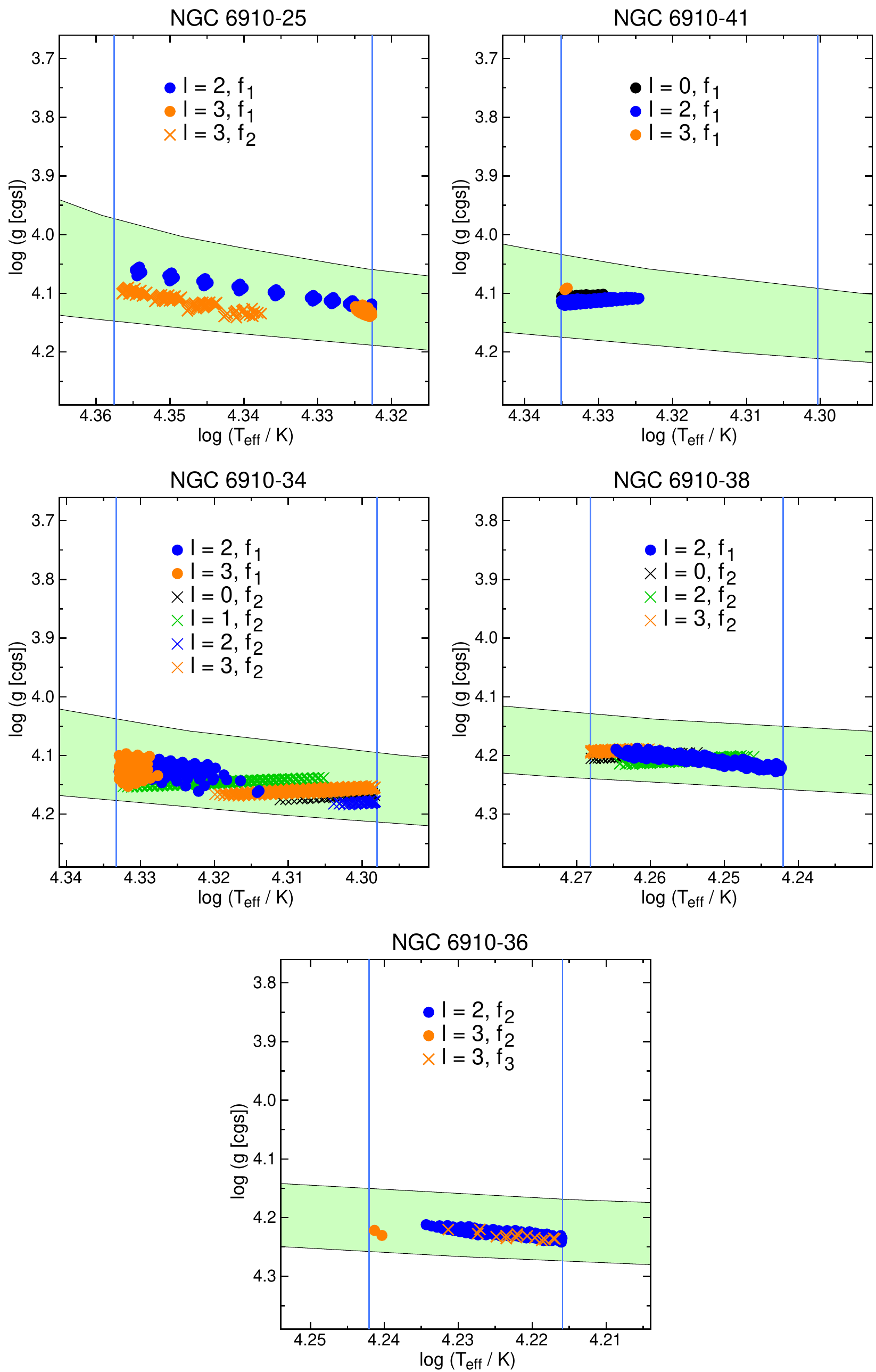}
\caption{Kiel diagrams for models of NGC\,6910-25, 41, 34, 38, and 36. The green area is the same as in Fig.\,\ref{Kiel0}. Blue vertical lines correspond to the adopted $T_{\rm eff}$ limits from Table \ref{atmpartab}. For these program stars, pulsational models for only a single rotational velocity $V_{\rm eq}=170$~{\kms} were considered. Symbols represent models that reproduce the labelled frequencies for a given $l$.}
\label{Kiel-low-mass}
\end{figure}

Two modes were found in NGC\,6910-25, one ($f_1$) has amplitude larger than 1~mmag. Only modes with $l=2$ and 3 reproduce this frequency for models in the allowable region. For $f_2$ only modes with $l=3$ fulfil the criteria, but the corresponding models do not overlap with those for $f_1$ (Fig.\,\ref{Kiel-low-mass}). A single mode has been found in NGC\,6910-41. It is identified as $l=2$, 0 or 3 (Fig.\,\ref{Kiel-low-mass}). If the mode is radial, it is a fundamental radial pulsation. 

The next two stars, NGC\,6910-34 and 38, show very similar frequency spectra with two frequencies of which one is $\sim$2.5 higher than the other. The lower frequencies ($f_1$ for both stars) are relatively high (3.8 and 4.8~d$^{-1}$), which would suggest that they are $p\text{}$ modes. However, both stars have relatively low masses for $\beta$~Cep stars and are located close to the zeor-age main sequence (ZAMS) (Fig.\,\ref{Kiel0}). The modes that reproduce their $f_1$ frequencies are therefore $g$ modes, while both $f_2$ are $p$ -modes. For both NGC\,6910-34 and 38, $f_2$ modes are identified by a requirement that models that reproduce $f_2$ overlap with those for $f_1$ in the Kiel diagram. NGC\,6910-38 is the coolest of the $\beta$~Cep stars in the cluster.

Three modes were detected in the coolest program star, NGC\,6910-36. The strongest, $f_1$, has a frequency 0.363~d$^{-1}$ and is clearly a $g$ mode. Two other modes, $f_2$ and $f_3$, are identified as $l=2$ and $l=3$, respectively (Fig.\,\ref{Kiel-low-mass}). Both are unambiguously identified as $g$ modes as well. NGC\,6910-36 is therefore a slowly pulsating B-type (SPB) star, which exhibits $g$ modes with relatively high frequencies, 5.19 and 4.73~d$^{-1}$.

\subsubsection{Second iteration of EnsA}\label{step-second}
After limiting the areas in the Kiel diagram in which the allowable models are located, we returned to the question whether these areas further limit the age range we adopted in Sect.\,\ref{step-age}. We found that this was the case. The limitations came from two stars, NGC\,6910-14 and 41. The allowable models for single-mode NGC\,6910-41 were not younger than 9.81~Myr, which changed the lower limit for the cluster age by 0.56~Myr. The allowable models for NGC\,6910-14 changed the upper limit to 11.48~Myr. 

With this new age range (9.81\,--\,11.48~Myr), we repeated the procedure of ensemble identification of $l$. This allowed us to exclude three identifications of $l$ marked with italics in Table \ref{modeidtab}. One of this changes leaves $f_7$ in NGC\,6910-18 identified unambiguously as $l=1$ mode. Thus, we completed the EnsA with an unambiguous ensemble identification of $l$ for eight modes. No further narrowing of cluster age was found, so that the final age of NGC\,6910 from the EnsA is 10.6$^{+0.9}_{-0.8}$~Myr. 

The classification of modes provided in Table \ref{modeidtab} enables us to conclude that eight of the nine program stars are $\beta$~Cep stars. The ninth star, NGC\,6910-36, is an SPB star because it shows only $g$ modes. In three of the eight $\beta$~Cep stars in the cluster (NGC\,6910-27, 34, and 38) we found a $g$ mode, which makes these stars hybrid $\beta$~Cep/SPB stars. With the lowering of the detection threshold in observations, especially using data from space, different studies show that hybridity is very common among $\beta$~Cep stars \citep{2011MNRAS.413.2403B,2016A&A...588A..55P,2018pas8.conf...64D,2019MNRAS.485.3544W}. This is confirmed by recent theoretical calculations, in which the extent of the $g$-mode instability domain is large in the upper part of the main sequence \citep{2015A&A...580L...9W,2016MNRAS.455L..67M,2017MNRAS.469...13S}.

\subsection{Parameters of the program stars}\label{params}
The final step of the EnsA is determining the stellar parameters for individual stars using evolutionary and seismic models. In general, the parameters ($T_{\rm eff}$, $\log g$, mass, $\alpha_{\rm ov}$, fraction of the main-sequence lifetime, and fraction of the mass of the stellar core) and their uncertainties were derived from a subset of evolutionary models that were used to obtain the ensemble $l$. In other words, these were the models that satisfied all five criteria we defined at the beginning of Sect.\,\ref{step-mid} for the strongest ($f_1$) mode in all program stars except for NGC\,6910-16, 27, and 36. For NGC\,6910-16 and 27 we used the region in which models overlapped that reproduce two modes, $f_1$ and $f_2$ (Fig.\,\ref{Kiel-14-16-27}), while for NGC\,6910-36 we took the region in which models reproduced $f_2$ with $l=2$ (Fig.\,\ref{Kiel-low-mass}). Finally, for NGC\,6910-18 the sum of the regions covered by models that reproduced $f_1$ as F and 1O mode (Fig.\,\ref{Kiel-18}) were used. The resulting parameters are given in Table \ref{eapartab}. For this star we also provide parameters for each of the two identifications separately. When its $\log g$ is determined with a precision better than $\sim$0.05~dex, it will be possible to conclude on the radial order of this mode. The value of $\log g$ presented in Table \ref{atmpartab} favours F, but uncertainties related to the determination of $\log g$ do not allow us to conclude that this is the correct identification. The values of the parameters are simple averages calculated for the subset of evolutionary models defined above; the asymmetric uncertainties reflect the span of a given parameter in the subset.

Table \ref{eapartab} also provides the fraction of the main-sequence lifetime, $f_{\rm MS}$, calculated as the ratio of the cluster age and the main-sequence lifetime for a given mass. It was calculated for models with $V_{\rm eq}=170$\,km\,s$^{-1}$ and $\alpha_{\rm ov}=0.2$. The $\beta$~Cep stars in NGC\,6910 cover a wide range of main-sequence evolution, starting from about 14\% of the main-sequence lifetime for NGC\,6910-38 up to over 80\% for the most massive program stars.
\begin{table*}
\centering\footnotesize
\caption{Parameters of the program stars obtained with ensemble asteroseismology taking into account $V_{\rm eq}$ listed in the second column. $f_{\rm MS}$ stands for the fraction of the main-sequence lifetime, and $M_{\rm c}/M$ for the fractional mass of the stellar core.}        
\label{eapartab}
\begin{tabular}{cccccccc} 
\hline\hline\noalign{\smallskip}
Star & $V_{\rm eq}$ [{\kms}] & \multicolumn{1}{c}{$T_{\rm eff}$ [K]} &  \multicolumn{1}{c}{$\log g$ (cgs)}  &  \multicolumn{1}{c}{Mass [$\rm M_\odot$]}  & $\alpha_{\rm ov}$ & $f_{\rm MS}$ & \multicolumn{1}{c}{$M_{\rm c}/M$}\\
\noalign{\smallskip}\hline\noalign{\smallskip}
14  & 170        & $26400^{+790}_{-400}$ & $3.714^{+0.233}_{-0.212}$ & $14.58^{+2.02}_{-2.38}$   & [0.1, 0.4]  & 0.82 & $0.287^{+0.020}_{-0.042}$\\\noalign{\smallskip}
16  & 170        & $25800^{+1090}_{-900}$ & $3.676^{+0.025}_{-0.020}$ & $14.54^{+0.86}_{-0.74}$  & [0.0, 0.4] & 0.81 & $0.257^{+0.047}_{-0.046}$\\\noalign{\smallskip}
27  & 170        & $25350^{+750}_{-1240}$ & $3.827^{+0.191}_{-0.131}$ & $12.75^{+1.85}_{-2.35}$  & [0.0, 0.4] & 0.67 & $0.255^{+0.043}_{-0.046}$\\\noalign{\smallskip}
18 (F +1O) & 97 -- 250 & $25030^{+470}_{-940}$ & $3.815^{+0.098}_{-0.095}$ & $12.50^{+1.20}_{-1.10}$ & [0.0, 0.4] & 0.65 & $0.244^{+0.049}_{-0.036}$\\\noalign{\smallskip}
   18 (F)     & 97 -- 250 & $24980^{+520}_{-890}$ & $3.877^{+0.036}_{-0.023}$ & $11.91^{+0.39}_{-0.51}$  & [0.0, 0.4] & 0.62 & $0.253^{+0.041}_{-0.029}$\\\noalign{\smallskip}
   18 (1O)    & 97 -- 250 & $25090^{+390}_{-750}$ & $3.740^{+0.030}_{-0.019}$ & $13.23^{+0.47}_{-0.53}$  & [0.0, 0.3] & 0.72 & $0.234^{+0.043}_{-0.026}$\\\noalign{\smallskip}
25  & 170        & $21620^{+1010}_{-600}$  & $4.102^{+0.035}_{-0.045}$ & $8.18^{+0.82}_{-0.52}$  & [0.0, 0.1] & 0.31 & $0.253^{+0.007}_{-0.009}$\\\noalign{\smallskip}
41  & 170        & $21450^{+180}_{-300}$  & $4.112^{+0.008}_{-0.020}$ & $8.01^{+0.19}_{-0.17}$   & [0.0, 0.3] & 0.30 & $0.253^{+0.013}_{-0.007}$\\\noalign{\smallskip}
34  & 170        & $21240^{+280}_{-640}$  & $4.125^{+0.038}_{-0.028}$ & $7.80^{+0.32}_{-0.56}$   & [0.0, 0.4] & 0.28 & $0.260^{+0.014}_{-0.014}$\\\noalign{\smallskip}
38  & 170        & $17890^{+500}_{-420}$  & $4.207^{+0.018}_{-0.018}$ & $5.63^{+0.33}_{-0.27}$   & [0.0, 0.4] & 0.14 & $0.251^{+0.007}_{-0.006}$\\\noalign{\smallskip}
36  & 170        & $16730^{+420}_{-290}$  & $4.226^{+0.015}_{-0.014}$ & $5.01^{+0.25}_{-0.19}$   & [0.0, 0.4] & 0.11 & $0.244^{+0.006}_{-0.005}$\\\noalign{\smallskip}  
\hline                  
\end{tabular}
\end{table*}

\begin{figure}[!t]
\centering
\includegraphics[width=\columnwidth]{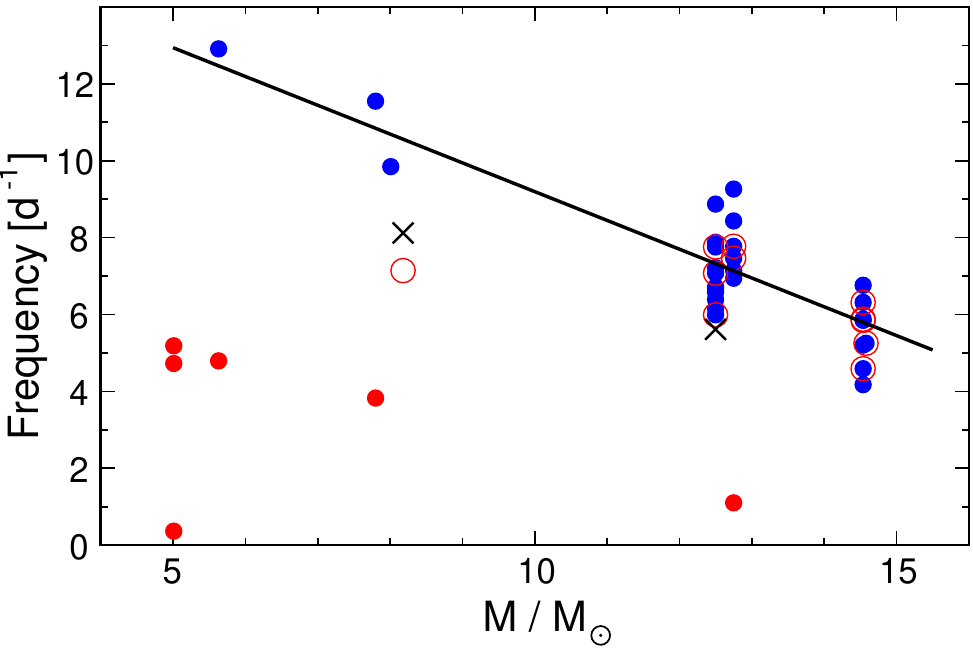}
\caption{Mass-frequency relation for program stars in NGC\,6910. Blue dots are $p$ modes, red dots are $g$ modes, encircled blue dots stand for modes classified as `$p$ or $p/g$' (Table \ref{modeidtab}), and the red circle is the mixed $p/g$ mode in NGC\,6910-25. Finally, two crosses are non-classified modes. The line is the relation given by \cite{Bal1997} fitted to $p$ modes; it corresponds to the age of 8.4~Myr.}
\label{mass-freq}
\end{figure}

The masses derived for the program stars allow us to plot frequencies as a function of mass in Fig.\,\ref{mass-freq}. This figure shows that $p$ and $g$ modes are clearly separated in this diagram. In addition, the sequence for $p$ modes, suggested from Fig.\,\ref{9fs}, is even better visible in Fig.\,\ref{mass-freq}. The relation shows also that frequency alone is not sufficient to distinguish between $p$ and $g$ modes. In the frequency range between 4\,d$^{-1}$ and 6\,d$^{-1}$ we find both $p$ and $g$ modes. This type of diagram has been used by \cite{Bal1997} to derive the ages of three southern clusters that are rich in $\beta$~Cep stars. A fit of frequencies of pure $p$ modes to the relation given by these authors gives an age of 8.4~Myr. This is lower than but within $\sim$3$\sigma$ uncertainty of the value obtained from the EnsA and is likely a result of different evolutionary models used by \cite{Bal1997}.

\subsection{Mode stability}\label{stability}
\begin{figure*}
\centering
\includegraphics[width=0.979\textwidth]{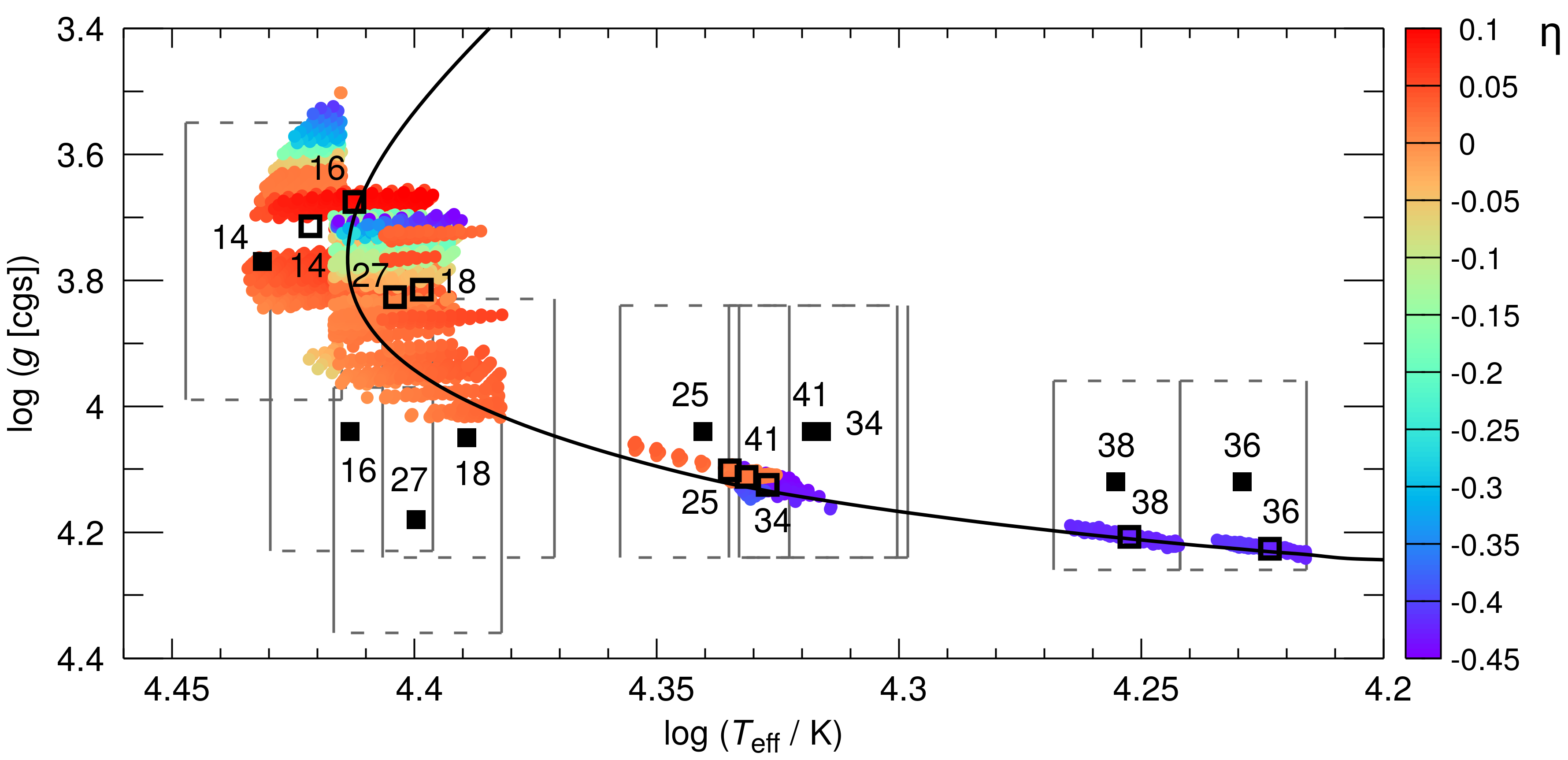}
\caption{Joint Kiel diagram for nine program stars in NGC\,6910. Only models comprising the five criteria defined at the beginning of Sect.\,\ref{step-mid} are plotted as colour dots, with the colour corresponding to the value of $\eta$ (right side bar). The filled and open squares denote the positions of the program stars according to the adopted parameters from Table \ref{atmpartab} and to the final determination from the EnsA (Table \ref{eapartab}), respectively. The black line is the 10.6-Myr isochrone calculated for $V_{\rm eq} = 170$\,km\,s$^{-1}$ and $\alpha_{\rm ov} = 0.2$.}
\label{BCep_puls}%
\end{figure*}

In the above procedure of EnsA we avoided discussing the stability of the identified modes except for a general requirement that the accepted modes must have $\eta>-0.5$ (Sect.\,\ref{step-mid}). Now, with the final results of the ensemble identification of $l$ (Table \ref{modeidtab}), it is time to return to this problem. Figure \ref{BCep_puls} shows the positions of all models we used for ensemble identification of $l$. In general there is no problem with the excitation of $p$ modes for the most massive program stars. The problem begins in less massive stars,  starting from NGC\,6910-41, the star with a mass of about 8.0\,M$_\odot$, $\eta$ becomes negative and the modes are formally stable. The two least massive $\beta$~Cep stars in the cluster, NGC\,6910-34 and 38, have masses of 7.8 and 5.6\,M$_\odot$ (Table \ref{eapartab}). For the $p$ modes in these stars, which have frequencies of 11.55\,d$^{-1}$ in NGC\,6910-34 and 12.91\,d$^{-1}$ in NGC\,6910-38, $\eta$~ranges between $-0.01$ and $-0.07$ for the former and between $-0.23$ and $-0.16$ for the latter star, depending on $l$ and the model. This means that even though the theory predicts that these modes are stable, they are not far from excitation. This is fairly consistent with the theoretical calculations of \cite{2015A&A...580L...9W}, in which the lower limit for $p$-mode instability corresponds to stars with masses of about 6 and 6.5\,M$_\odot$ for models with OPLIB and OPAL opacities, respectively. A small discrepancy between the observations and theory remains, however.

The situation is worse for $g$ modes. Figure\,\ref{BCep_puls} shows that the observed $g$ modes in NGC\,6910-34, 38, and 36 are stable. This is in general a problem that prompts a revision of the opacities that are used to calculate seismic models. The standard opacities cannot explain the excitation of low-frequency $g$ modes in hybrid stars \citep{2018pas8.conf...64D}. The present study shows that a similar problem also arises for $p$ modes and $g$ modes with high ($>3.5$\,d$^{-1}$) frequencies in stars with masses of 5\,--\,8~M$_\odot$. The discrepancy seems to prompt a revision of opacities, but the recipe of how to change them cannot be concluded from our work.

\section{Summary and discussion}\label{discussion}
As a result of the ground-based photometric and spectroscopic campaign, nine stars with at least one frequency higher than 5~d$^{-1}$, candidates for $\beta$~Cep-type stars, were found in the young open cluster NGC\,6910. The main purpose of this work was to make use of the constraints resulting from the membership of these stars, primarily the age of the cluster, to perform ensemble asteroseismology. The procedure started with the classical photometric and spectroscopic mode identification, which resulted in an unambiguous identification of degree $l$ for only five of 37 modes, including the  radial mode in NGC\,6910-18. The other constraints were (i) the age of the cluster limited owing to the radial mode in NGC\,6910-18 and the single mode found in the most massive of the program stars, NGC\,6910-14, and (ii) the atmospheric parameters, in particular $T_{\rm eff}$, derived from spectra and/or Str\"omgren photometry (Table \ref{atmpartab}). The final result of the EnsA, consisting of two iterations of modelling individual stars (Sect.\,\ref{step-mid}), is a cluster age of 10.6$^{+0.9}_{-0.8}$~Myr, ensemble identification of eight modes, and their classification as $p$, $g$, or mixed $p/g$ modes (Table \ref{modeidtab}). In consequence, we conclude that of the nine program stars, eight are $\beta$~Cep stars (including three $\beta$~Cep/SPB hybrids) while the least massive star, NGC\,6910-36, is an SPB star. Finally, we obtained stellar parameters for all program stars (Table \ref{eapartab}). The precise values of $\log g$ derived by EnsA for the less massive program stars are particularly promising. For the more massive stars (e.g.~for NGC\,6910-18 and 27) the final $\log g$ is significantly different from the values derived in Sect.\,\ref{AtmPar_ch} from spectra and Str\"omgren photometry. In some cases (NGC\,6910-41, 34) we were able to narrow $T_{\rm eff}$ even when the modes were not identified unambiguously, like in NGC\,6910-41.

All this shows that EnsA is a fruitful method that can be successfully used in modelling pulsating stars in open clusters. The results we present here are obviously model dependent. However, the other methods for determining cluster ages, the isochrone fitting in particular, are model dependent as well. For example, \cite{Zibi2004} derived the age of NGC\,6910 by means of isochrone fitting for 6\,$\pm$\,2~Myr, which is significantly lower than in the present paper. They used the isochrones of \cite{1994A&AS..106..275B} that are available in terms of absolute magnitude and colour. These isochrones are transformed from the theoretical ($\log L/L_\odot$, $\log T_{\rm eff}$) to observational (absolute magnitude, colour) Hertzsprung-Russell diagram using calibrations that are subject to uncertainties, especially related to bolometric corrections. In addition, the isochrone fitting suffers from the correction for reddening, which is frequently variable across clusters with ages younger than $\sim$20~Myr because the remnant matter of the parent cloud did not have enough time to disperse. Differential reddening is rarely corrected prior to isochrone fitting. The other difficulties are the scarcity of stars near the turn-off point, the high binary fraction among the most massive members, and fast rotation, which causes magnitudes and colours of stars to depend on the (generally unknown) inclination of the rotational axis \citep[e.g.][]{2018AJ....156..165C}. As a result, for open clusters with ages comparable to that of NGC\,6910 ($\sim$10\,Myr), the precision of the age determination via isochrone fitting is rarely better than 50\%, even when good-quality photometry is used. NGC\,6910 itself is a good illustration of this problem. The ages of this cluster obtained with isochrone fitting range between about 6~Myr \citep{2000AJ....119.1848D,Zibi2004} to 21~Myr \citep{2005A&A...438.1163K}. In this context, a $\sim$10\% precision of the age determination obtained by means of EnsA for NGC\,6910 and $\sim$20\% for $\chi$~Per \citep{Saes2013} seems to be a significant improvement in age determination.

A method that might provide a comparably precise determination of the age of a young open cluster is based on a strong dependence of stellar radius on age during the main-sequence evolution of a star with a mass higher than $\sim$15\,M$_\odot$. For the method to work, both mass and radius have to be derived precisely, which means that it is applicable only to massive components of eclipsing binaries. An example of using this method was presented by \cite{2013MNRAS.429.1354M} for ALS\,1135 in the Bochum\,7 association. Not many stellar systems contain massive eclipsing binaries like this, however.

The results of the EnsA can be used for testing seismic models. As indicated in Sect.\,\ref{stability}, not all observed modes are found to be unstable in seismic models. This is the case for some $p$ modes, but mostly for the $g$ modes in NGC\,6910-34, 38, and 36 (Fig.\,\ref{BCep_puls}). The possible revisions of the opacities can therefore benefit from the EnsA.  Our considerations do not result in strong limitations for $\alpha_{\rm ov}$; only for four program stars did we obtain some limits for $\alpha_{\rm ov}$  (Table \ref{eapartab}).

The procedure of EnsA can be supplemented by adding constraints from modelling member eclipsing binaries as this could allow for an independent determination of their masses. In this way, evolutionary models could be tested as a part of EnsA. The brightest eclipsing binary in NGC\,6910 is NGC\,6910-30 ($V=12.9$~mag). Only three eclipses were partly covered by the campaign data, hence we can only conclude that the orbital period was longer than 40\,d and the orbit was eccentric. The inclusion of this star in the EnsA would require a dedicated photometric and spectroscopic campaign on the cluster, however.

Based on the results we obtained for NGC\,6910, the prospects for applying EnsA to other young open clusters seems promising. One of the most obvious targets is the other target of the 2005\,--\,2007 campaign, NGC\,869 (h Persei). This cluster is already known to be rich in $\beta$~Cep stars \citep{2008JPhCS.118a2068M} and contains bright eclipsing binaries such as Oo\,1516 ($V=11.0$~mag), as reported by the same authors.

\begin{acknowledgements}
Based on observations obtained with the Apache Point Observatory 3.5-meter telescope, which is owned and operated by the Astrophysical Research Consortium, Nordic Optical Telescope, operated by the Nordic Optical Telescope Scientific Association at the Observatorio del Roque de los Muchachos, La Palma, Spain, of the Instituto de Astrof\'{\i}sica de Canarias, and Mercator Telescope, operated on the island of La Palma by the Flemish Community, at the Spanish Observatorio del Roque de los Muchachos of the Instituto de Astrof\'{\i}sica de Canarias. We thank Eva Bauwens, Bart Vandenbussche, Alexander Eigenbrod, Christoffel Waelkens, Pieter Deroo, Erik Broeders, Djazia Ladjal, Wim De Meester, Cezary Ku{\l}akowski, Evelien Vanhollebeke, Rik Huygen, Rachel Drummond, Roy {\O}stensen, Matthieu Karrer, Elena Puga Antol\'{\i}n, Laurent Le Guillou, and Rosa Mar\'{\i}a Dom\'{\i}nguez Quintero for making some observations of NGC\,6910. This work was supported by the NCN grants 2012/05/N/ST9/03898 and 2016/21/B/ST9/01126 and has received funding from the European Community’s Seventh Framework Programme (FP7/2007-2013) under grant agreement no.~269194. PW acknowledges support from NCN grants 2013/08/S/ST9/00583 and 2015/17/B/ST9/02082. JNF acknowledges the support from the National Natural Science Foundation of China (NSFC) through the grants 11833002 and 11673003. TM acknowledges financial support from the European Space Agency through a Postdoctoral Research Fellow grant and from the Research Council of Leuven University through grant GOA/2003/04. CA and CG receive funding from the European Research Council (ERC) under the European Union’s Horizon 2020 research and innovation programme (grant agreement N$^\circ$670519: MAMSIE). This research has made use of the WEBDA database, operated at the Department of Theoretical Physics and Astrophysics of the Masaryk University.
\end{acknowledgements}
\bibliographystyle{aa}
\bibliography{EnsA_NGC6910_v1.7}
%-------------------------------------------------------------------
\begin{appendix}%First online appendix
\section{Parameters of the sine-curve fitting to the $U_{\rm G}BVI_{\rm C}$ light curves of program stars in NGC\,6910.}\label{sinefits}
\longtab[1]{
\centering\footnotesize
\begin{landscape}
\begin{longtable}{crrrrrrr@{.}lr@{.}lr@{.}lr@{.}l}
\caption{\label{1stPerPar}The parameters of the sine-curve fitting to the $U_{\rm G}BVI_{\rm C}$ light curves of program stars in NGC\,6910.}\\
\hline\hline\noalign{\smallskip}
\multicolumn{1}{c}{Star}    & \multicolumn{2}{c}{Frequency} & \multicolumn{4}{c}{Amplitude [mmag]} & \multicolumn{8}{c}{$T_{\rm max}$ (HJD -- 2453500.0)}\\
\multicolumn{1}{c}{(WEBDA)} & \multicolumn{2}{c}{[d$^{-1}$]}     & \multicolumn{1}{c}{$U_{\rm G}$} & \multicolumn{1}{c}{$B$} & \multicolumn{1}{c}{$V$} &\multicolumn{1}{c}{$I_{\rm C}$} & \multicolumn{2}{c}{$U_{\rm G}$} & \multicolumn{2}{c}{$B$} & \multicolumn{2}{c}{$V$} & \multicolumn{2}{c}{$I_{\rm C}$} \\
\noalign{\smallskip}\hline\noalign{\smallskip}
\endfirsthead
\caption{continued.}\\
\hline\hline\noalign{\smallskip}
\multicolumn{1}{c}{Star}    & \multicolumn{2}{c}{Frequency} & \multicolumn{4}{c}{Amplitude [mmag]} & \multicolumn{8}{c}{$T_{\rm max}$ (HJD -- 2453500.0)}\\
\multicolumn{1}{c}{(WEBDA)} & \multicolumn{2}{c}{[d$^{-1}$]}     & \multicolumn{1}{c}{$U_{\rm G}$} & \multicolumn{1}{c}{$B$} & \multicolumn{1}{c}{$V$} &\multicolumn{1}{c}{$I_{\rm C}$} & \multicolumn{2}{c}{$U_{\rm G}$} & \multicolumn{2}{c}{$B$} & \multicolumn{2}{c}{$V$} & \multicolumn{2}{c}{$I_{\rm C}$} \\
\noalign{\smallskip}\hline\noalign{\smallskip}
\endhead
\hline
\endfoot
\hline
\endlastfoot
14 & $f_1$ &  5.252056 & 8.44(13) & 8.41(09) & 8.52(04) & 8.61(06) & 379&3433(05) & 514&14651(34) & 418&37591(13) & 499&86707(22)\\
 & $ 2\times f_1$ & 10.504112 &  & *0.40(09) & 0.30(04) & 0.46(06) & \multicolumn{2}{c}{---} & *514&1962(35) & 418&3269(19) & 499&8195(21)\\
\noalign{\smallskip}\hline\noalign{\smallskip}
16 & $f_1$ & 5.202740 & 12.29(17) & 10.83(09) & 10.49(04) & 9.67(06) & 389&9114(04) & 514&07702(26) & 408&74717(13) & 498&31610(18)\\
 & $f_2$ & 4.174670 & 6.47(17) & 5.91(09) & 5.39(04) & 5.20(06) & 389&8328(10) & 514&1584(06) & 408&75917(31) & 498&3475(4)\\
 & $f_3$ & 5.846318 & 5.02(17) & 4.21(10) & 4.20(04) & 4.71(06) & 389&9295(09) & 514&1110(06) & 408&74670(30) & 498&2042(04)\\
 & $f_4$ & 4.588471 & 2.22(16) & 1.79(09) & 1.96(04) & 1.58(06) & 389&9942(26) & 514&2095(18) & 408&7333(08) & 498&3034(13)\\
 & $f_5$ & 5.878679 & 1.28(17) & 1.73(10) & 1.92(04) & 1.74(06) & 389&8358(35) & 514&1945(15) & 408&7267(07) & 498&2025(10)\\
 & $f_6$ & 6.759432 &  ---  & 0.89(09) & 0.82(04) & 0.87(06) & \multicolumn{2}{c}{---} & 514&0846(25) & 408&7498(13) & 498&2542(16)\\
 & $ 2\times f_2$ & 8.349340 &  ---  & 0.45(09) & 0.69(04) & 0.62(06) & \multicolumn{2}{c}{---} & 514&1523(40) & 408&7564(12) & 498&2250(18)\\
 & $f_7$ & 6.314395 &  ---  & 0.80(09) & 0.62(04) & *0.45(06) & \multicolumn{2}{c}{---} & 514&0365(30) & 408&7160(18) & *498&1957(33)\\
 & $f_1+f_3$ & 11.049058 &  ---  & 0.54(09) & 0.49(04) & 0.57(06) & \multicolumn{2}{c}{---} & 514&1378(25) & 408&7904(13) & 498&2977(15)\\
\noalign{\smallskip}\hline\noalign{\smallskip} 
27 & $f_1$ & 6.942973 & 6.62(22) & 5.38(11) & 4.50(05) & 4.40(07) & 378&4208(08) & 513&9523(05) & 417&74036(24) & 498&9738(04)\\
 & $f_2$ & 7.773690 & 3.04(23) & 3.40(11) & 3.14(05) & 3.48(07) & 378&3202(15) & 513&9032(07) & 417&6815(03) & 498&8536(04)\\
 & $f_3$ & 1.100920 & --- & 1.85(11) & 1.78(05) & 1.59(06) & \multicolumn{2}{c}{---} & 513&6583(89) & 417&3634(38) & 499&1638(62)\\
 & $f_4$ & 8.433329 & 1.66(23) & 1.57(11) & 1.53(05) & 1.57(07) & 378&3075(26) & 513&9557(14) & 417&6698(06) & 498&8969(08)\\
 & $f_5$ & 7.463826 & 0.94(24) & 1.11(11) & 1.19(05) & 1.24(07) & 378&3670(51) & 513&9538(22) & 417&7565(09) & 498&9483(12)\\
 & $f_6$ & 7.146300 & --- & 0.86(11) & 0.98(05) & 0.89(07) & \multicolumn{2}{c}{---} & 513&9128(29) & 417&6347(11) & 498&9397(17)\\
 & $f_7$ & 9.262165 & --- & 0.59(11) & 0.69(05) & *0.36(07) &\multicolumn{2}{c}{---}  & 513&8658(33) & 417&6698(12) & *498&9680(32)\\
\noalign{\smallskip}\hline\noalign{\smallskip}
18 & $f_1$ & 6.154885 & 17.70(21) & 10.47(09) & 9.54(04) & 7.40(06) & 393&51144(30) & 513&90543(23) & 418&20938(10) & 498&30943(20)\\
 & $f_2$ & 6.388421 & 9.20(20) & 8.48(09) & 8.23(04) & 7.63(06) & 393&5620(06) & 513&93681(27) & 418&13855(11) & 498&28343(19)\\
 & $f_3$ & 6.715615 & 5.28(21) & 4.85(09) & 4.73(04) & 4.37(06) & 393&6350(09) & 513&9564(05) & 418&05966(19) & 498&32093(33)\\
 & $f_4$ & 5.991368 &  ---  & 0.90(09) & 1.47(04) & 1.10(06) &  \multicolumn{2}{c}{---}& 513&9751(27) & 418&1794(07) & 498&2944(14)\\
 & $f_5$ & 8.871393 &  ---  & 1.68(09) & 1.43(04) & 1.37(06) &  \multicolumn{2}{c}{---}& 513&9014(10) & 418&0893(05) & 498&3471(08)\\
 & $f_6$ & 7.074318 &  ---  & 1.60(09) & 1.40(04) & 1.05(06) &  \multicolumn{2}{c}{---}& 513&9628(13) & 418&1278(06) & 498&2697(13)\\
 & $f_7$ & 6.658287 &  ---  & 1.23(09) & 1.41(04) & 1.45(06) &  \multicolumn{2}{c}{---}& 513&8822(18) & 418&0605(07) & 498&2641(10)\\
 & $f_8$ & 7.879447 &  ---  & 1.37(09) & 1.22(04) & 1.24(06) &  \multicolumn{2}{c}{---}& 513&8933(14) & 418&0788(07) & 498&2835(10)\\
 & $f_9$ & 7.755512 &  ---  & 0.87(09) & 0.94(04) & 0.97(06) &  \multicolumn{2}{c}{---}& 513&8991(22) & 418&0915(09) & 498&2972(12)\\
 & $f_{10}$ & 7.219347 & ---   & 0.88(09) & 0.75(04) & 0.97(06) &  \multicolumn{2}{c}{---}& 513&9531(23) & 418&0963(11) & 498&2988(13)\\
 & $f_{11}$ & 5.612781 &  ---  & 0.80(09) & 0.85(04) & 0.60(06) &  \multicolumn{2}{c}{---}& 513&9123(04) & 418&0544(13) & 498&2344(28)\\
 & $f_{12}$ & 6.579093 &  ---  & 0.84(09) & 0.72(04) & 0.53(06) &  \multicolumn{2}{c}{---}& 513&9068(27) & 418&1499(13) & 498&2559(27)\\
\noalign{\smallskip}\hline\noalign{\smallskip}
25 & $f_1$ & 7.141122 & 1.69(23) & 1.53(14) & 1.72(06) & 1.15(11) & 381&8310(30) & 513&7388(20) & 397&3704(07) & 498&0528(21)\\
 & $f_2$ & 8.120426 & --- & 1.06(14) & 1.01(05) & *0.69(11) &  \multicolumn{2}{c}{---} & 513&7860(26) & 397&2946(11) & *498&0183(31)\\
\noalign{\smallskip}\hline\noalign{\smallskip}
41 & $f_1$ & 9.845570 & 5.21(42) & 3.28(21) & 3.14(08) & 2.47(08) & 373&5395(14) & 514&5172(11) & 419&4474(04) & 498&3664(06)\\
\noalign{\smallskip}\hline\noalign{\smallskip}
34 & $f_1$ & 3.826208 & 2.57(33) & 2.77(12) & 2.76(06) & 2.15(09) & 370&6916(55) & 514&1820(18) & 349&5260(09) & 498&7645(18)\\
 & $f_2$ & 11.548531 & --- & 0.75(12) & 0.64(06) & 0.70(09) & \multicolumn{2}{c}{---} & 514&2377(22) & 349&5427(13) & 498&7368(18)\\
\noalign{\smallskip}\hline\noalign{\smallskip}
38 & $f_1$ & 4.793720 & --- & 2.23(20) & 2.26(08) & 2.21(11) &\multicolumn{2}{c}{---}  & 514&1806(30) & 350&0066(12) & 497&9088(17)\\
 & $f_2$ & 12.909208 & --- & 1.05(20) & 0.85(08) & 0.69(11) & \multicolumn{2}{c}{---} & 514&11112(23) & 349&9621(12) & 497&9165(20)\\
\noalign{\smallskip}\hline\noalign{\smallskip}
36 & $f_1$ & 0.363151 & 3.62(43) & 2.97(18) & 3.69(07) & 3.20(11) & 384&018(44) & 513&409(26) & 416&959(08) & 496&906(15)\\
 & $f_2$ & 5.185733 & ---  & 1.92(18) & 1.87(07) & 2.02(11) & \multicolumn{2}{c}{---} & 514&3289(29) & 418&1123(11) & 497&1720(17)\\
 & $f_3$ & 4.725566 & --- & 1.47(18) & 1.59(07) & 1.28(11) & \multicolumn{2}{c}{---} & 514&3030(41) & 418&0108(14) & 497&1654(29)\\
\end{longtable}
\tablefoot{` * '  preceding amplitude and phase denotes no mode detection in the light curve in this filter.}
\end{landscape}
}% End longtab

\longtab[2]{
\centering\footnotesize
\begin{landscape}
\begin{longtable}{ccrrrrrr@{.}lr@{.}lr@{.}l}
\caption{\label{2ndPerPar}Same as in Table \ref{1stPerPar} but for the 2013 observations.}\\
\hline\hline\noalign{\smallskip}
Star & $V$ & \multicolumn{2}{c}{Frequency} & \multicolumn{3}{c}{Amplitude [mmag]} & \multicolumn{6}{c}{$T_{\rm max}$ (HJD -- 2456300.0)}\\
(WEBDA) & [mag] & \multicolumn{2}{c}{[d$^{-1}$]} & \multicolumn{1}{c}{$B$} & \multicolumn{1}{c}{$V$} &\multicolumn{1}{c}{$I_{\rm C}$} & \multicolumn{2}{c}{$B$} & \multicolumn{2}{c}{$V$} & \multicolumn{2}{c}{$I_{\rm C}$}\\
\noalign{\smallskip}\hline\noalign{\smallskip}
\endfirsthead
\caption{continued.}\\
\hline\hline\noalign{\smallskip}
Star & $V$ & \multicolumn{2}{c}{Frequency} & \multicolumn{3}{c}{Amplitude [mmag]} & \multicolumn{6}{c}{$T_{\rm max}$ (HJD -- 2456300.0)}\\
(WEBDA) & [mag] & \multicolumn{2}{c}{[d$^{-1}$]} & \multicolumn{1}{c}{$B$} & \multicolumn{1}{c}{$V$} &\multicolumn{1}{c}{$I_{\rm C}$} & \multicolumn{2}{c}{$B$} & \multicolumn{2}{c}{$V$} & \multicolumn{2}{c}{$I_{\rm C}$}\\
\noalign{\smallskip}\hline\noalign{\smallskip}
\endhead
\hline
\endfoot
\hline
\endlastfoot
14 & 10.432 & $f_1$ & 5.25232 & 9.12(25) & 9.52(12) & 9.06(25) & 218&4828(35) & 225&9094(26) & 221&149(06)\\
16 & 10.626 & $f_1$ & 5.20251 & 15.58(35) & 15.30(16) & 14.73(27) & 218&3222(07) & 226&0116(22) & 221&5871(06)\\
 &  & $f_2$ & 5.84568 & 6.55(35) & 6.30(15) & 7.16(27) & 218&2123(15) & 225&9097(50) & 221&6306(10)\\
 &  & $f_3$ & 4.17542 & 4.65(34) & 4.37(15) & 3.80(26) & 218&2246(29) & 225&884(10) & 221&5758(27)\\
27 & 11.735 & $f_1$ & 6.94376 & 4.10(50) & 3.73(21) & *2.63(35) & 217&1056(28) & 225&313(10) & *221&2849(31)\\
 &  & $f_2$ & 7.46595 & 3.45(51) & 3.40(21) & 2.29(36) & 217&2012(31) & 225&371(11) & 221&3599(33)\\
18 & 10.849 & $f_1$ & 6.15422 & 10.08(36) & 9.40(16) & 6.84(35) & 218&2396(46) & 225&7157(32) & 221&1657(13)\\
 &  & $f_2$ & 6.38858 & 8.81(35) & 8.08(15) & 7.18(34) & 218&1552(10) & 225&6663(34) & 221&1284(12)\\
 &  & $f_3$ & 6.71532 & 4.68(37) & 4.37(16) & 3.94(35) & 218&2526(19) & 225&6948(64) & 221&2322(21)\\
25 & 11.497 & $f_1$ & 7.14313 & 1.46(29) & 1.98(14) & *1.54(30) & 218&410(05) & 223&306(12) & *221&204(05)\\
41 & 12.873 & $f_1$ & 9.84548 & 3.18(62) & 2.50(24) & *2.25(45) & 217&6883(32) & 225&2047(16) & *221&5453(32)\\
34 & 12.752 & $f_1$ & 3.82833 & 2.94(37) & *2.09(15) & *2.28(29) & 218&0203(34) & *225&8520(30) & *221&683(05)\\
36 & 13.050 & $f_1$ & 0.36128 & 3.85(37) & 4.04(14) & *4.33(33) & 218&918(44) & 224&39(12) & *221&803(34)\\
\end{longtable}
\tablefoot{` * ' sign preceding amplitude and phase denotes no mode detection in this filter light curve.}
\end{landscape}
}% End longtab

\section{Radial velocites of NGC\,6910-14 and 18.}\label{RVs-14-18}
\begin{table}[!h]
\caption{Radial velocites of NGC\,6910-14.}
\label{RVs-14}
\centering\small
\begin{tabular}{ccccccccc}
\hline\hline\noalign{\smallskip}
BJD$_{\rm TDB}$ & RV & $\sigma_{\rm RV}$&BJD$_{\rm TDB}$ & RV & $\sigma_{\rm RV}$&BJD$_{\rm TDB}$ & RV & $\sigma_{\rm RV}$\\
$-2453500.0$ & [{\kms}]  & [{\kms}] &$-2453500.0$ & [{\kms}]  & [{\kms}] &$-2453500.0$ & [{\kms}]  & [{\kms}] \\
\noalign{\smallskip}\hline\noalign{\smallskip}
869.357279 & $-$27.38 & 2.19 & 887.321448 & $-$26.67 & 0.64 & 890.346187 & $-$25.67 & 0.92 \\
869.379235 & $-$28.02 & 1.70 & 887.334134 & $-$24.65 & 0.64 & 891.242275 & $-$27.56 & 1.08 \\
869.398239 & $-$30.03 & 0.80 & 887.342869 & $-$23.00 & 0.62 & 891.257182 & $-$27.87 & 1.39 \\
869.491800 & $-$24.24 & 2.39 & 887.350822 & $-$22.35 & 0.55 & 891.271396 & $-$28.79 & 1.22 \\
869.511649 & $-$26.13 & 2.11 & 888.234074 & $-$25.08 & 1.20 & 891.285610 & $-$28.57 & 1.38 \\
869.530352 & $-$27.37 & 1.86 & 888.253827 & $-$24.98 & 0.70 & 891.300346 & $-$30.34 & 0.87 \\
869.561486 & $-$28.40 & 1.31 & 888.264167 & $-$24.08 & 0.74 & 891.314943 & $-$28.34 & 0.52 \\
869.577376 & $-$29.35 & 0.77 & 888.273749 & $-$23.59 & 0.64 & 891.341445 & $-$26.66 & 0.34 \\
870.340245 & $-$28.98 & 1.61 & 888.283327 & $-$24.45 & 0.71 & 892.243388 & $-$28.19 & 0.48 \\
870.357200 & $-$28.72 & 1.92 & 888.293506 & $-$22.59 & 0.65 & 892.258470 & $-$25.61 & 0.62 \\
870.372015 & $-$27.35 & 1.76 & 888.303080 & $-$22.33 & 0.78 & 892.273147 & $-$25.22 & 0.41 \\
870.388345 & $-$28.14 & 1.82 & 888.312651 & $-$23.66 & 0.84 & 892.287823 & $-$22.46 & 0.66 \\
870.416272 & $-$25.92 & 0.38 & 888.322755 & $-$22.19 & 0.72 & 892.302408 & $-$24.80 & 0.73 \\
870.424976 & $-$24.89 & 1.86 & 888.332354 & $-$21.20 & 0.63 & 892.316995 & $-$23.49 & 0.42 \\
870.482659 & $-$28.53 & 2.19 & 888.341936 & $-$20.66 & 0.51 & 892.331752 & $-$22.40 & 0.42 \\
871.355234 & $-$26.81 & 2.60 & 889.242013 & $-$23.14 & 1.11 & 893.242179 & $-$25.65 & 1.17 \\
871.369562 & $-$25.54 & 3.20 & 889.257268 & $-$21.73 & 0.94 & 893.256795 & $-$23.56 & 0.92 \\
871.382119 & $-$25.65 & 2.27 & 889.271983 & $-$22.35 & 1.20 & 893.272237 & $-$22.56 & 0.83 \\
871.396147 & $-$25.32 & 2.64 & 889.286628 & $-$23.24 & 1.05 & 893.287094 & $-$21.92 & 0.83 \\
871.412315 & $-$27.22 & 3.15 & 889.301277 & $-$25.10 & 0.97 & 893.301666 & $-$21.72 & 0.81 \\
871.424815 & $-$28.02 & 2.13 & 889.316166 & $-$25.16 & 1.27 & 893.316525 & $-$22.71 & 0.79 \\
887.250543 & $-$27.95 & 1.12 & 889.330380 & $-$26.97 & 0.96 & 893.331242 & $-$24.38 & 0.78 \\
887.263091 & $-$25.87 & 0.73 & 889.344605 & $-$28.47 & 1.03 & 894.249084 & $-$22.93 & 0.47 \\
887.271924 & $-$28.41 & 0.49 & 890.259524 & $-$24.21 & 1.15 & 894.263713 & $-$24.05 & 0.44 \\
887.281067 & $-$27.16 & 0.70 & 890.274143 & $-$25.10 & 1.03 & 894.278409 & $-$24.93 & 0.45 \\
887.289018 & $-$27.98 & 0.81 & 890.288363 & $-$25.22 & 1.18 & 894.293294 & $-$26.71 & 0.66 \\
887.296975 & $-$25.90 & 0.94 & 890.302578 & $-$25.38 & 1.30 & 894.307512 & $-$27.02 & 0.66 \\
887.305487 & $-$26.34 & 0.71 & 890.317755 & $-$27.21 & 0.82 & 894.322216 & $-$26.74 & 0.64 \\
887.313485 & $-$27.39 & 0.60 & 890.331972 & $-$26.46 & 1.00 &&&\\
\noalign{\smallskip}\hline
\end{tabular}
\end{table}

\begin{table}[!h]
\caption{Radial velocites of NGC\,6910-18.}
\label{RVs-18}
\centering\small
\begin{tabular}{rccrccrcc}
\hline\hline\noalign{\smallskip}
BJD$_{\rm TDB}$ & RV & $\sigma_{\rm RV}$&BJD$_{\rm TDB}$ & RV & $\sigma_{\rm RV}$&BJD$_{\rm TDB}$ & RV & $\sigma_{\rm RV}$\\
$-2453500.0$ & [{\kms}]  & [{\kms}] &$-2453500.0$ & [{\kms}]  & [{\kms}] &$-2453500.0$ & [{\kms}]  & [{\kms}] \\
\noalign{\smallskip}\hline\noalign{\smallskip}
868.398281 & $-$18.03 & 0.41 & 3037.826897 & $-$29.91 & 0.37 & 3064.663593 & $-$29.12 & 0.31 \\
868.425571 & $-$22.30 & 0.44 & 3037.836999 & $-$26.40 & 0.37 & 3064.673687 & $-$30.67 & 0.29 \\
868.447318 & $-$23.59 & 0.49 & 3037.848691 & $-$25.81 & 0.43 & 3064.683781 & $-$31.87 & 0.29 \\
868.469088 & $-$22.96 & 0.38 & 3037.864561 & $-$26.44 & 0.52 & 3064.694196 & $-$33.78 & 0.32 \\
868.492560 & $-$21.18 & 0.44 & 3037.878414 & $-$30.01 & 0.54 & 3064.704290 & $-$36.40 & 0.35 \\
868.514318 & $-$20.18 & 0.35 & 3037.893167 & $-$36.17 & 0.58 & 3064.716721 & $-$38.27 & 0.29 \\
868.536065 & $-$18.15 & 0.42 & 3037.912341 & $-$42.63 & 0.48 & 3064.726815 & $-$39.78 & 0.30 \\
868.557998 & $-$18.48 & 0.44 & 3064.577488 & $-$38.50 & 0.36 & 3064.736909 & $-$41.28 & 0.34 \\
868.581203 & $-$18.70 & 0.42 & 3064.587582 & $-$37.59 & 0.26 & 3064.747003 & $-$41.28 & 0.41 \\
2986.821664 & $-$26.94 & 0.28 & 3064.597676 & $-$36.46 & 0.30 & 3064.757097 & $-$39.72 & 0.29 \\
2986.887578 & $-$40.58 & 0.29 & 3064.610421 & $-$36.05 & 0.39 & 3064.779678 & $-$33.80 & 0.27 \\
3018.818079 & $-$34.70 & 0.28 & 3064.620525 & $-$34.04 & 0.28 & 3064.789793 & $-$32.83 & 0.25 \\
3018.966346 & $-$36.50 & 0.34 & 3064.630620 & $-$32.37 & 0.53 & 3064.798326 & $-$31.13 & 0.38 \\
3037.781069 & $-$41.98 & 0.33 & 3064.640713 & $-$31.05 & 0.29 &&&\\
3037.791163 & $-$40.76 & 0.41 & 3064.650808 & $-$30.24 & 0.34 &&&\\
\noalign{\smallskip}\hline
\end{tabular}
\end{table}

\section{Diagnostic $\chi_{\rm T}^2$ diagrams for identifications of the spherical harmonic degree $l$.}\label{Xt_diagrams}%Second appendix
\begin{figure*}[!h]
\centering
\includegraphics[width=\textwidth]{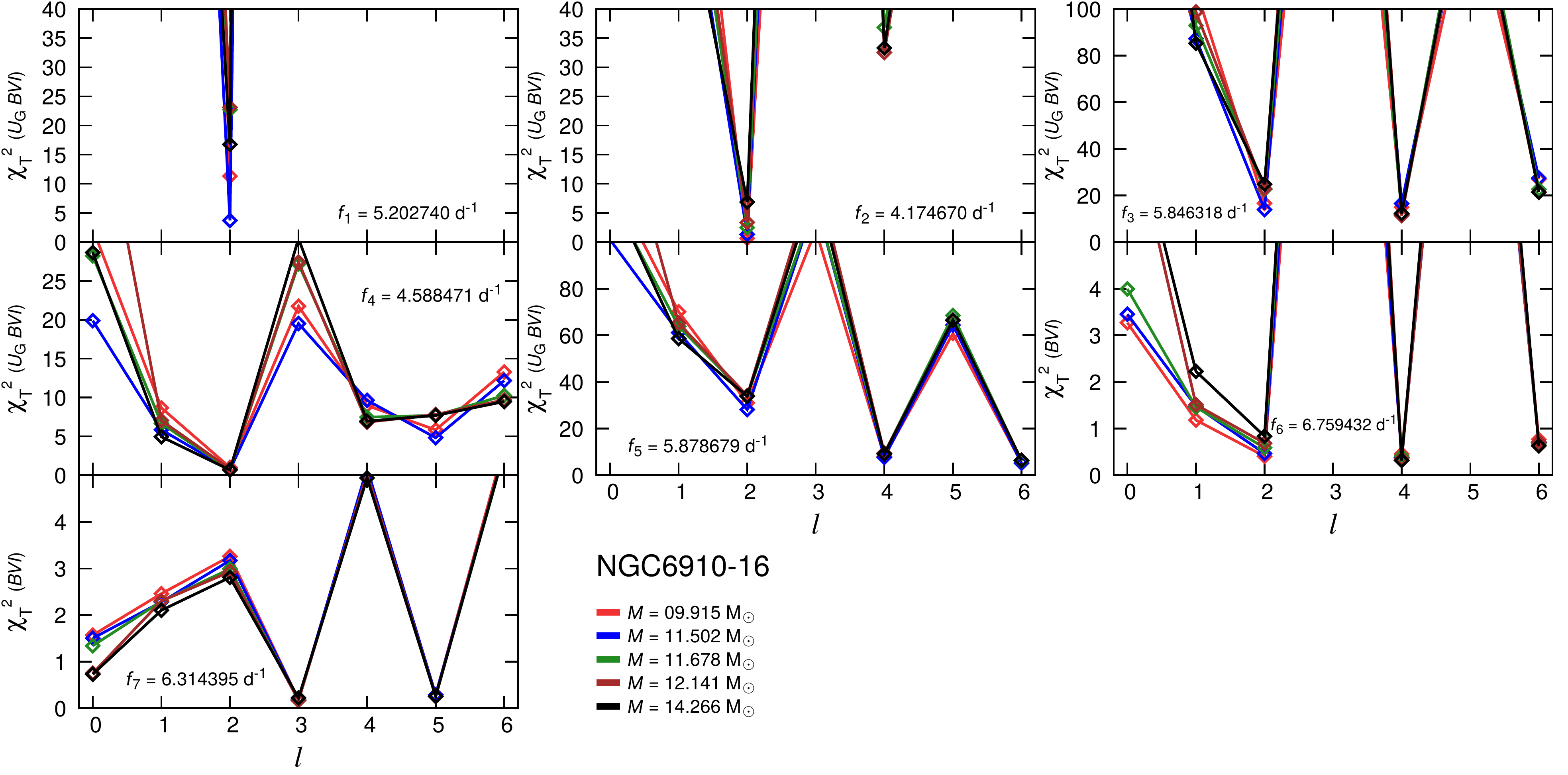}
\caption {Photometric mode identification by means of the $\chi^2_{\rm T}$ parameter for seven modes detected in NGC\,6910-16. The filters listed in parentheses denote which amplitudes were used to derive $\chi^2_{\rm T}$. Different colours stand for different models (labeled).}
\label{16_ampident}
\end{figure*}
\begin{figure*}[!h]
\centering
\includegraphics[width=\textwidth]{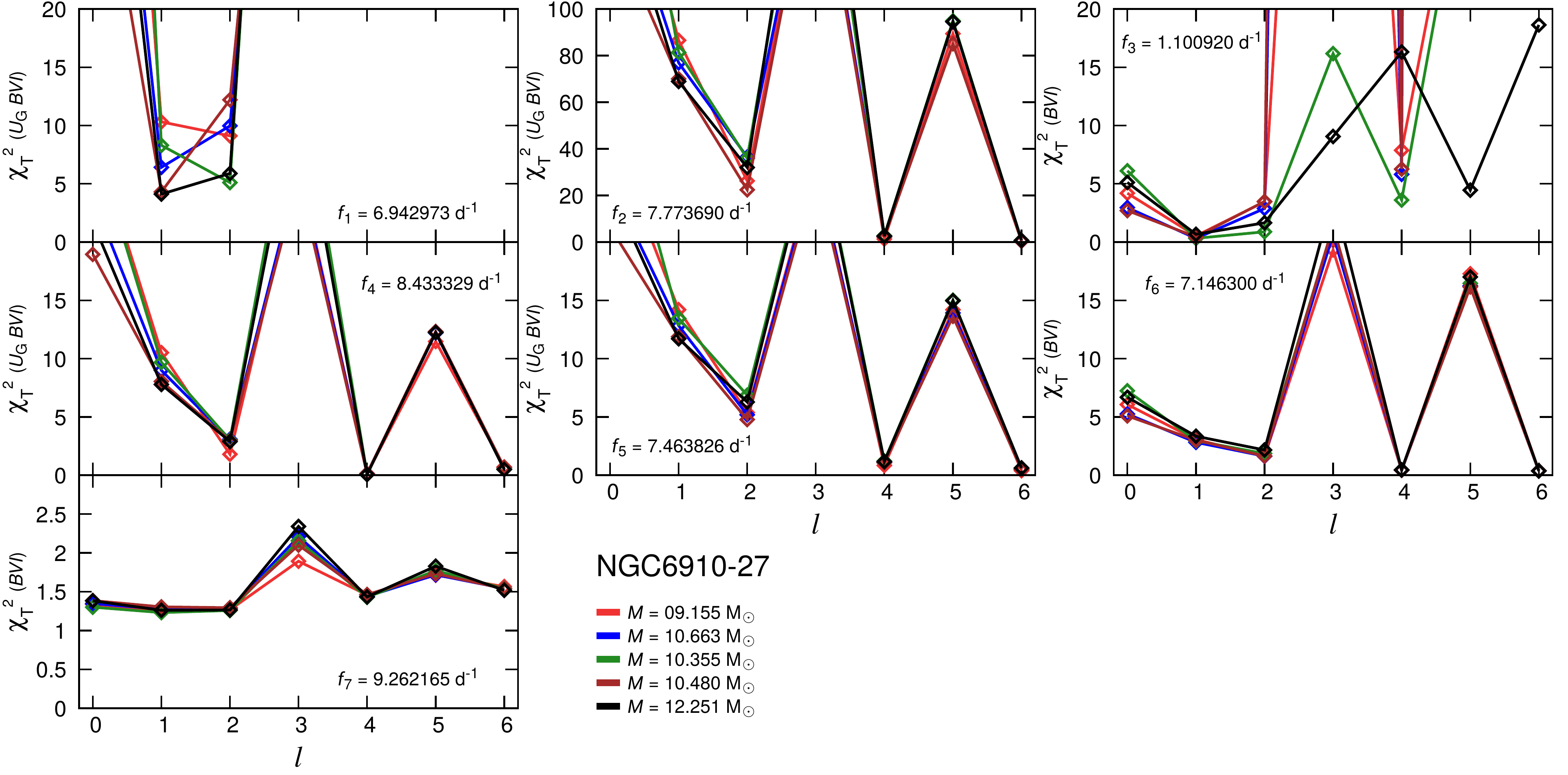}
\caption {Same as in Fig.\,\ref{16_ampident} but for seven modes detected in NGC\,6910-27.}
\label{27_ampident}
\end{figure*}
\begin{figure*}[!h]
\centering
\includegraphics[width=\textwidth]{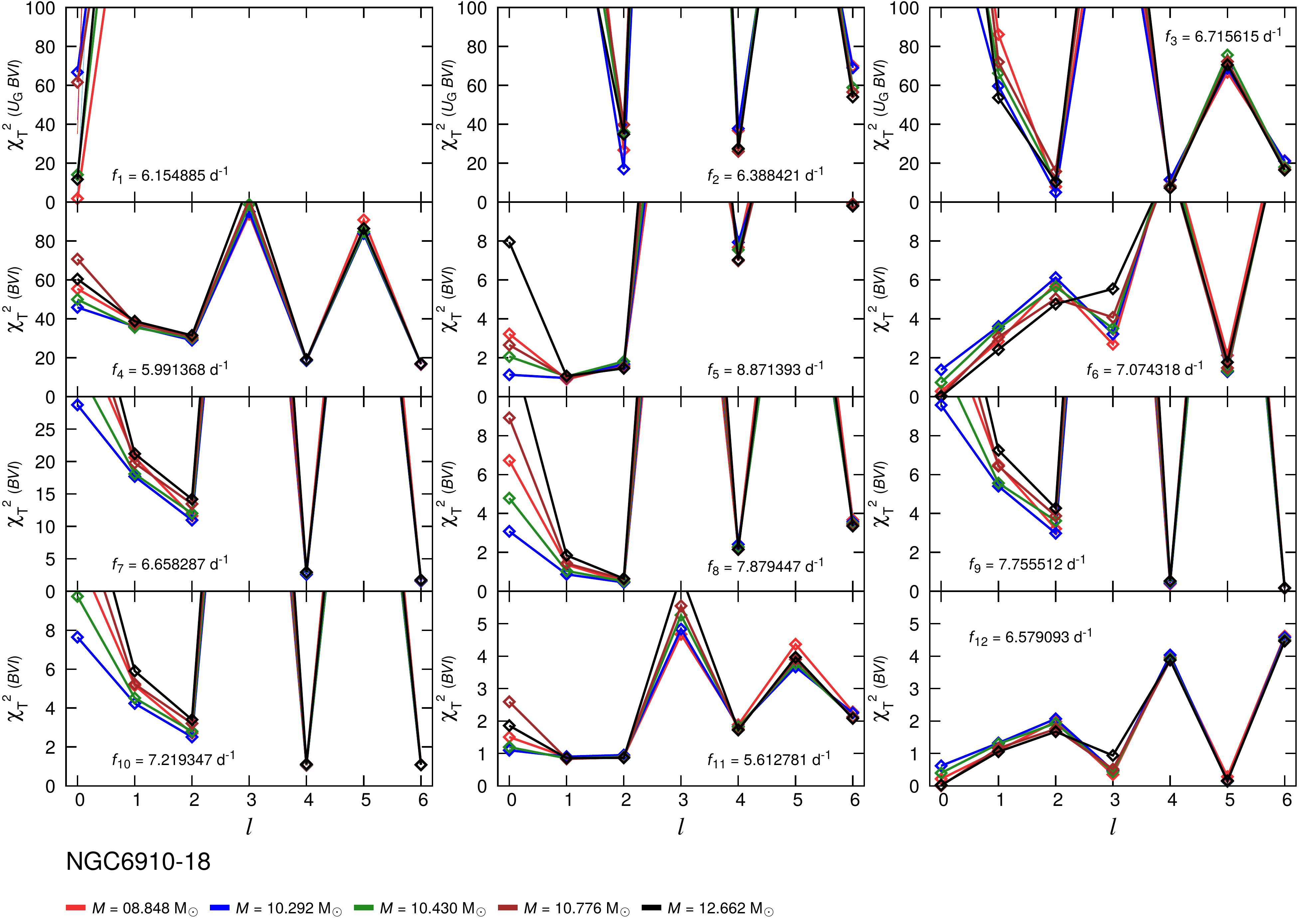}
\caption {Same as in Fig.\,\ref{16_ampident} but for twelve modes detected in NGC\,6910-18.}
\label{18_ampident}
\end{figure*}
\begin{figure*}[!h]
\centering
\includegraphics[width=\textwidth]{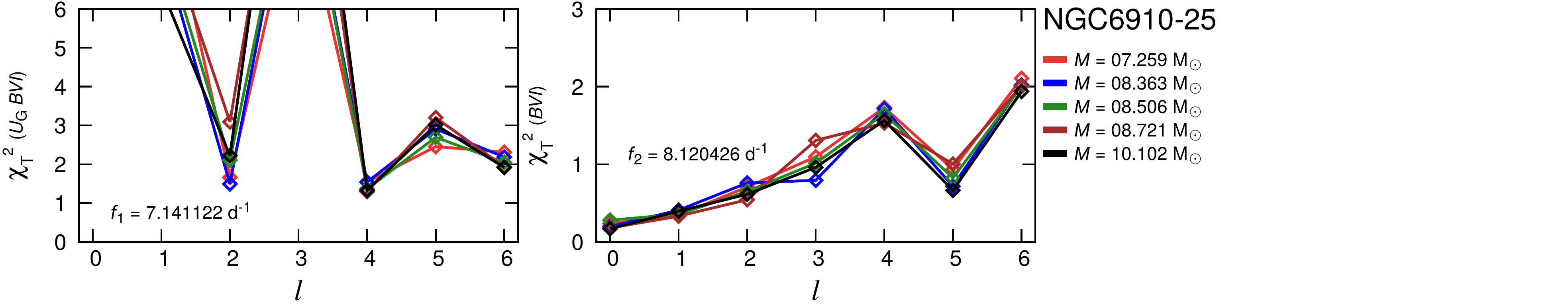}
\caption {Same as in Fig.\,\ref{16_ampident} but for two modes detected in NGC\,6910-25.}
\label{25_ampident}
\end{figure*}
\begin{figure*}[!h]
\centering
\includegraphics[width=\textwidth]{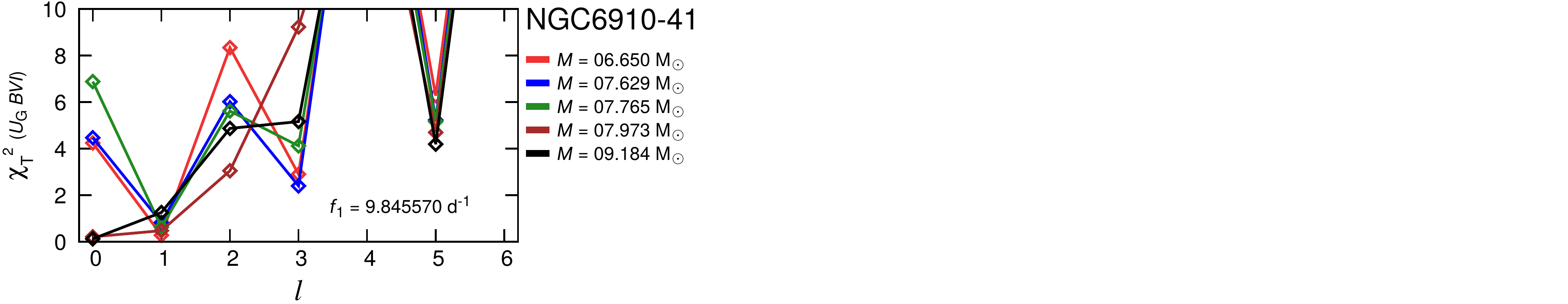}
\caption {Same as in Fig.\,\ref{16_ampident} but for the only mode detected in NGC\,6910-41.}
\label{41_ampident}
\end{figure*}
\begin{figure*}[!h]
\centering
\includegraphics[width=\textwidth]{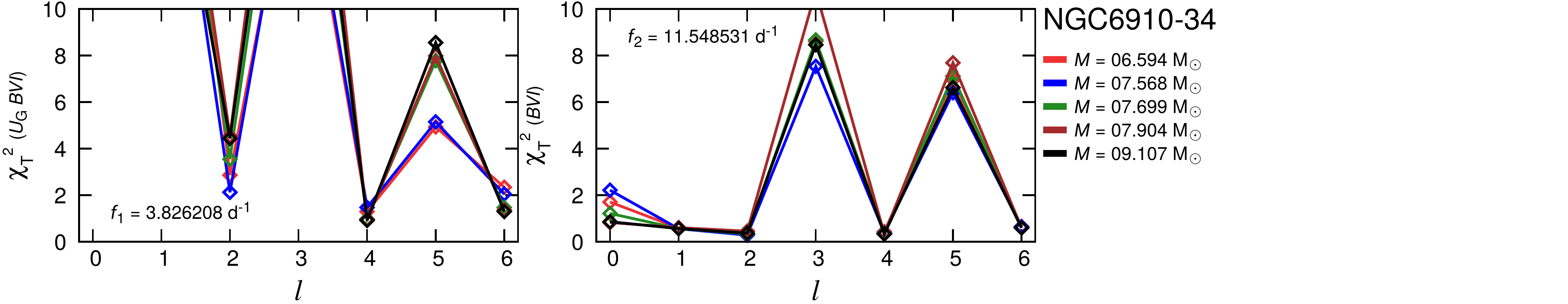}
\caption {Same as in Fig.\,\ref{16_ampident} but for two modes detected in NGC\,6910-34.}
\label{34_ampident}
\end{figure*}
\begin{figure*}[!h]
\centering
\includegraphics[width=\textwidth]{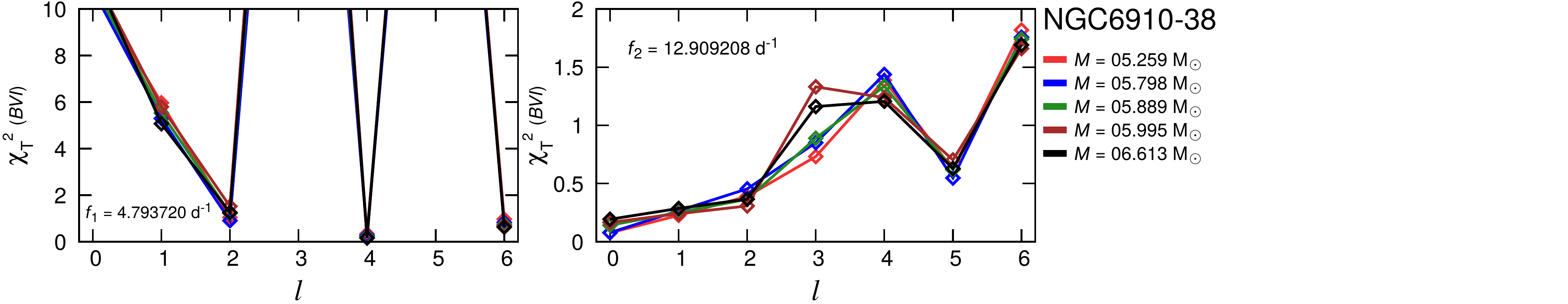}
\caption {Same as in Fig.\,\ref{16_ampident} but for two modes detected in NGC\,6910-38.}
\label{38_ampident}
\end{figure*}
\begin{figure*}[!h]
\centering
\includegraphics[width=\textwidth]{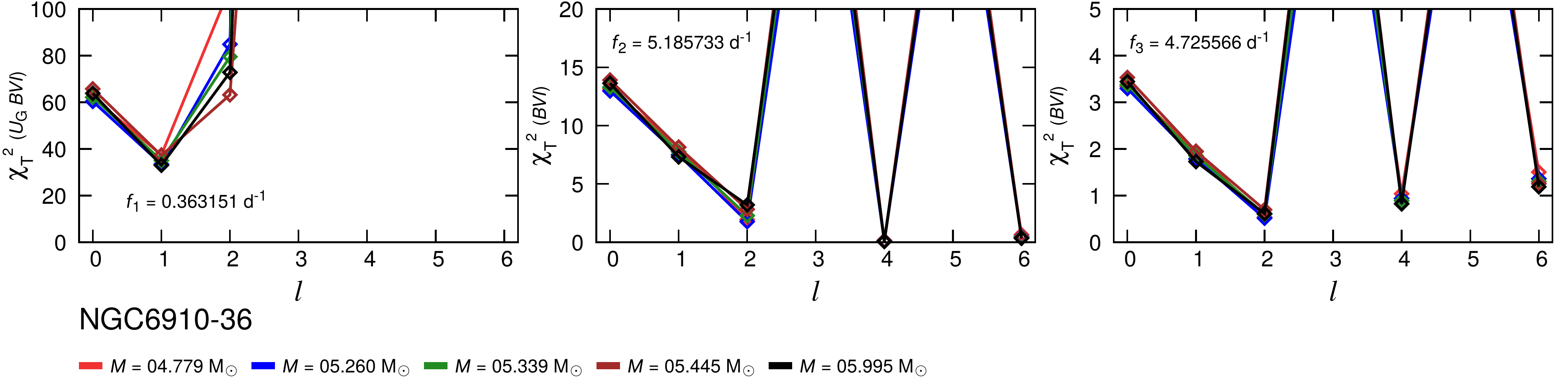}
\caption {Same as in Fig.\,\ref{16_ampident} but for three modes detected in NGC\,6910-36.}
\label{36_ampident}
\end{figure*}

\end{appendix}
\end{document}